\documentclass[pdflatex,sn-mathphys-num]{sn-jnl}

\usepackage{graphicx}%
\usepackage{multirow}%
\usepackage{amsmath,amssymb,amsfonts}%
\usepackage{amsthm}%
\usepackage{mathrsfs}%
\usepackage[title]{appendix}%
\usepackage{xcolor}%
\usepackage{textcomp}%
\usepackage{manyfoot}%
\usepackage{booktabs}%
\usepackage{algorithm}%
\usepackage{algorithmicx}%
\usepackage{algpseudocode}%
\usepackage{listings}%
\usepackage{placeins}

\usepackage{verbatim}

\theoremstyle{thmstyleone}%
\theoremstyle{thmstyletwo}%

\theoremstyle{thmstylethree}%

\definecolor{pinegreen}{rgb}{0.0, 0.47, 0.44}

\begin{document}

\title[Article Title]{GEN$^2$: A Generative Prediction-Correction Framework for Long-time Emulations of Spatially-Resolved Climate Extremes}


\author[1]{\fnm{Mengze} \sur{Wang}}
\equalcont{These authors contributed equally to this work.}

\author[1]{\fnm{Benedikt} \sur{Barthel Sorensen}}
\equalcont{These authors contributed equally to this work.}

\author*[1]{\fnm{Themistoklis P.} \sur{Sapsis}}\email{sapsis@mit.edu}

\affil*[1]{\orgdiv{Department of Mechanical Engineering}, \orgname{Massachusetts Institute of Technology}, \orgaddress{\city{Cambridge}, \postcode{02139}, \state{MA}, \country{USA}}}

\abstract{
Accurately quantifying the increased risks of climate extremes requires generating large ensembles of climate realization across a wide range of emissions scenarios, which is computationally challenging for conventional Earth System Models.
We propose GEN$^2$, a generative prediction-correction framework for an efficient and accurate forecast of the extreme event statistics.
The prediction step is constructed as a conditional Gaussian emulator, followed by a non-Gaussian machine-learning (ML) correction step.
The ML model is trained on pairs of the reference data and the emulated fields nudged towards the reference, to ensure the training is robust to chaos.
We first validate the accuracy of our model on historical ERA5 data and then demonstrate the extrapolation capabilities on various future climate change scenarios.
When trained on a single realization of one warming scenario, our model accurately predicts the statistics of extreme events in different scenarios, successfully extrapolating beyond the distribution of training data.
}




\maketitle


\section{Introduction}\label{sec:intro}
It is widely expected that the current rapid rate of climate change will lead to an increase in the frequency of extreme weather events such as tropical storm, heatwaves, and droughts \cite{raymond_understanding_2020,robinson_increasing_2021,fischer_increasing_2021}. These events can have massive negative impacts on society lost lives and economic costs which have already balooned from several million dollars in 1980 to 368 billion in 2024 \cite{allen_managing_2012,houser_economic_2015,AON2025report,fiedler_business_2021}. Quantifying the increased risk of these extreme events as a function of various climate change scenarios is the first step in providing policymakers with the information needed to implement both plan for and mitigate their impacts on society. However, despite their increasing frequency, extreme weather events remain rare and thus it is impossible to accurately quantify their frequency or severity purely from observations.
Accurate risk assessment requires generating sufficient data in the form of ensembles of long time horizon climate simulations. Furthermore, the trajectory of various driving factors, such as greenhouse gas emissions over the next several decades is far from clear, and depends on various social, political, economic, and technological factors we make no attempt to predict here. Therefore the amount of required data is further multiplied by the need to test a variety of emissions scenarios Earth may plausibly encounter. In summary, the challenge of quantifying extreme weather risks reduces to a need to generate large ensemble of climate realizations both across and within projected emissions scenarios.

The current state-of-the-art for generating such climate realizations relies on Earth System Models (ESMs) which solve the dynamical equations governing the earth's atmosphere, oceans, and biosphere.\cite{smagorinsky_general_1963,smagorinsky_numerical_1965,manabe_simulated_1965,mintz_very_1968,taylor2009non,dennis2012cam, golaz2022doe}. Relying on ESMs as a primary predictive tool presents two challenges. First, there is a vast array of physical processes and interactions affecting the atmosphere that we do not yet fully understand, and must therefore be empirically parametrized in our numerical models \cite{stensrud_parameterization_2007,holloway_moisture_2009,friend_carbon_2014,bloom_decadal_2016}. Second, practical computational cost restricts global simulations to a spatial resolution of approximately 100 km. Such a coarse resolution not only precludes the quantification of weather events evolving on smaller length scales, but also leads to inaccuracy in the larger resolved scales.
Many studies have sought to ameliorate these challenges through the introduction of data-driven forcing terms which are intended to parametrize the effects of any unknown physics and/or unresolved scales \cite{rasp_deep_2018,brenowitz_spatially_2019,yuval_use_2021,watt-meyer_correcting_2021,bretherton_correcting_2022,clark_correcting_2022}.
Despite such innovations, simulating the Earth System over a long time horizon on spatially-resolved grids pushes the frontier capabilities of modern high-performance computing.
This has motivated the need for cheaper data-driven reduced-order models to augment or even replace numerical ESMs.

One approach driven by the recent advances in machine learning are auto-regressive weather models, which aim to predict the atmospheric state at a certain time as a function of the state at previous times \cite{pathak_fourcastnet_2022,lam2023graphcast,bodnar2024Aurora,bi2023Pangu}. The accuracy of these models generally compares favorably to that of numerical weather predictions, but at a fraction of the cost.
However, numerical instabilities limit their predictions to a few days or potentially weeks.
Even if the instability issue can be resolved \cite{watt2023ace}, it is computationally expensive to run machine-learning weather models for hundreds of years to predict the climate.
As we are interested in the statistics of events occurring on multi-decade or longer time scales, we do not pursue this approach here.
An alternative approach is to construct reduced-complexity models for the climate system, or so-called climate emulators.
These models focused on quantifying the parametric relationships between various inputs to the climate system, such as greenhouse gas emissions, and the climate response.
One of the most widely used approaches is known as Linear Pattern Scaling (LPS), where the local climate variables are assumed to be linear functions of the global mean temperature \cite{Mitchell2003pattern,tebaldi_pattern_2014,osborn_performance_2018}.
The general LPS approach has recently been expanded in a variety of ways including accounting for physical processes such as emission history \cite{Castruccio2014trajectory,freese2024spatially} and internal variability \cite{Beusch2020MESMER,Link2019PCA} as well as the incorporation of more sophisticated metrics for modeling spatial correlation \cite{Alexeeff2018pattern}. Furthermore, various deep learning based alternatives to LPS have been proposed \cite{watson-parris_climatebench_2022,Kaltenborn2023ClimateSet}. However, \citet{lutjens_impact_2024} recently compared the performance of varioius emulators on ClimateBench \cite{Watson2022climatebench} and found the benefits of deep learning emulators are at best unclear. Additionally, emulation generally predicts time averaged quantities, and only a few recent studies have explored emulating extreme events such as the annual maximum temperature \cite{Quilcaille2022MESMERX} and heat wave duration \cite{Tebaldi2020extreme,Quilcaille2022MESMERX}.


Despite the success of LPS and its variants, the inherent nonlinearity and non-Gaussianity of the climate system places an upper bound on their potential to predict the full statistics of extremes, such as joint distributions of different variables. 
Correcting the biases of these emulators is known as \textit{debiasing}.
The most widely used strategy for debiasing coarse-resolution climate models is to augment numerical models with machine-learned parametric forcing terms, which aim to mimic the effects of the unresolved ``sub-grid scale'' dynamics \cite{rasp_deep_2018,brenowitz_spatially_2019,yuval_use_2021,watt-meyer_correcting_2021,bretherton_correcting_2022,clark_correcting_2022}.
However, like the fully auto-regressive models mentioned previously, these \textit{intrusive} approaches suffer from instabilities when integrated over long ($10+$ year) time horizons \cite{wikner_stabilizing_2022,yuval_use_2021}.
An alternative strategy is \textit{non-intrusive} debiasing, which corrects the output of imperfect models in a post-processing manner -- thereby bypassing the stability issue.
The challenge with non-intrusive debiasing is that learning a map between two arbitrary chaotic trajectories is generally ill-posed, and any such map will not generalize to unseen data during training. 
Learning a generalizable map requires \textit{paired} training data that are minimally affected by chaotic divergence.
This is possible through the framework introduced by \citet{barthel_sorensen_non-intrusive_2024}, which relies on training a correction operator on a surrogate model nudged towards a high fidelity reference.
By formulating the supervised learning problem directly between paired trajectories, this strategy facilitates learning the dynamics with very little training data, which in turn enables the extrapolation of statistics when the learned map is applied to much longer trajectories \cite{barthel_sorensen_non-intrusive_2024,barthel_sorensen_probabilistic_2024} and out-of-sample climate change scenarios \cite{zhang_machine_2024}.
However, these innovations still require expensive ESM simulations to generate the data needed for training and inference.
In this work we aim to replace these expensive computations with parsimonious climate emulators.

Our approach, which we refer to as ``GEN$^2$'' -- as it consists of two generative steps applied in succession:
(1) A Gaussian emulator that correctly captures the second-order spatio-temporal statistics of the climate,
(2) A diffusion-model-based debiasing step trained using the nudging framework introduced by \citet{barthel_sorensen_non-intrusive_2024}.
The emulation step is built on the stochastic model introduced by \citet{wang2023emulator}, which we have extended to emulate multiple variables, including wind speed, temperature, and humidity.
The emulator is also refined to capture the spatio-temporal spectra of the variable of interest.
The debiasing step is achieved using a conditional diffusion model \cite{song_score-based_2021,bischoff_unpaired_2023} whose architectural backbone is based on U-Net introduced by \citet{ronneberger_u-net_2015}.
The choice of diffusion model allows for significantly improved debiasing capabilities as compared to the simpler auto-encoder based models used in previous studies, as has been recently demonstrated on tasks including debiasing \cite{wan_debias_2023} and down-scaling \cite{lopez-gomez_dynamical-generative_2024,wan_statistical_2024}.
Additionally, it is an inherently probabilistic model meaning that a single input can be used to generate a distribution of outputs.
We first validate our approach on historical ERA5 data \cite{hersbach2020era5}, and then demonstrate its extrapolation capabilities on various climate change scenarios for the coming century.

\begin{figure}[h]
    \centering
    \includegraphics[width=0.95\textwidth]{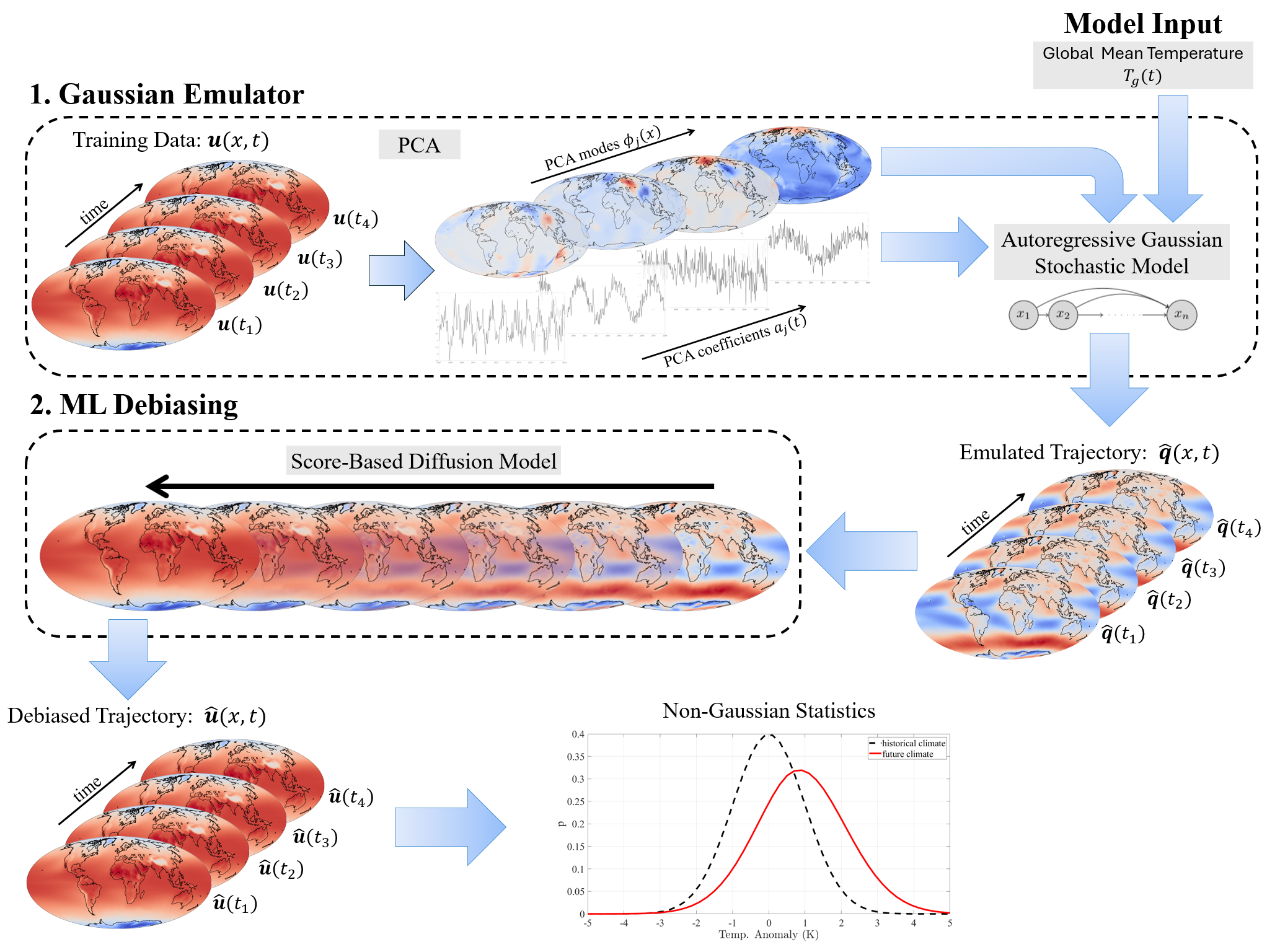}
    \caption{Schematic diagram of the proposed two-step GEN$^2$ emulation framework consisting of an initial Gaussian emulator followed by a ML correction. The model takes as an input a time series of global mean temperature: $T_g(t)$ and outputs climate trajectory defined over the entire globe $\hat{\boldsymbol{u}}(\boldsymbol{x},t)$. The Gaussian emulator assumes a spatial basis of PCA modes computed from the training data, and models the temporal coefficients as an autoregressive Gaussian process. The ML debiasing step consists of a score-based diffusion model trained to debias long term chaotic trajectories.  *The specific data-images shown are chosen purely for illustrative purposes, and the bias of the emulator is exaggerated.}
    \label{fig:method_diagram}
\end{figure}

\section{Results}\label{sec:results}
Our goal is to accurately quantify the spatial and temporal statistics of low probability, high impact weather events over long time horizons given solely the global mean temperature. As the latter is known to be proportional to cumulative $CO_2$ emissions, this allows us to directly quantify the spatial distribution of evolving extreme weather risk in a changing climate. We will use the term ``climate'' to refer to the statistics of the atmospheric state as quantified by the zonal and meridional wind speed: $U,V$, temperature: $T$, and specific humidity: $Q$.
The GEN$^2$ framework takes a prescribed global mean temperature trajectory $T_g(t)$ as input--a scalar quantity--and outputs the full field trajectory of prognostic variables defined on a global grid, whose resolution is determined by the specific training data used.
As illustrated in figure \ref{fig:method_diagram}, the model consists of two components: an initial Gaussian emulation step and a diffusion-based ML debiasing step. The Gaussian emulation derives its spatial structure from the training data and assumes temporal dynamics are Gaussian processes conditioned on the global mean temperature. The subsequent debiasing step aims to reconstruct the strongly non-Gaussian tail statistics of extreme weather events. Both components are trained on the same reference dataset, although this is not necessary in practice. For brevity, the following discussion focuses on a subset of variables and statistics, with additional results provided in the Supplemental Information (see Supplementary Notes). 

\subsection{Historical Validation}
\begin{figure}[h]
    \centering
    \includegraphics[width=0.95\textwidth]{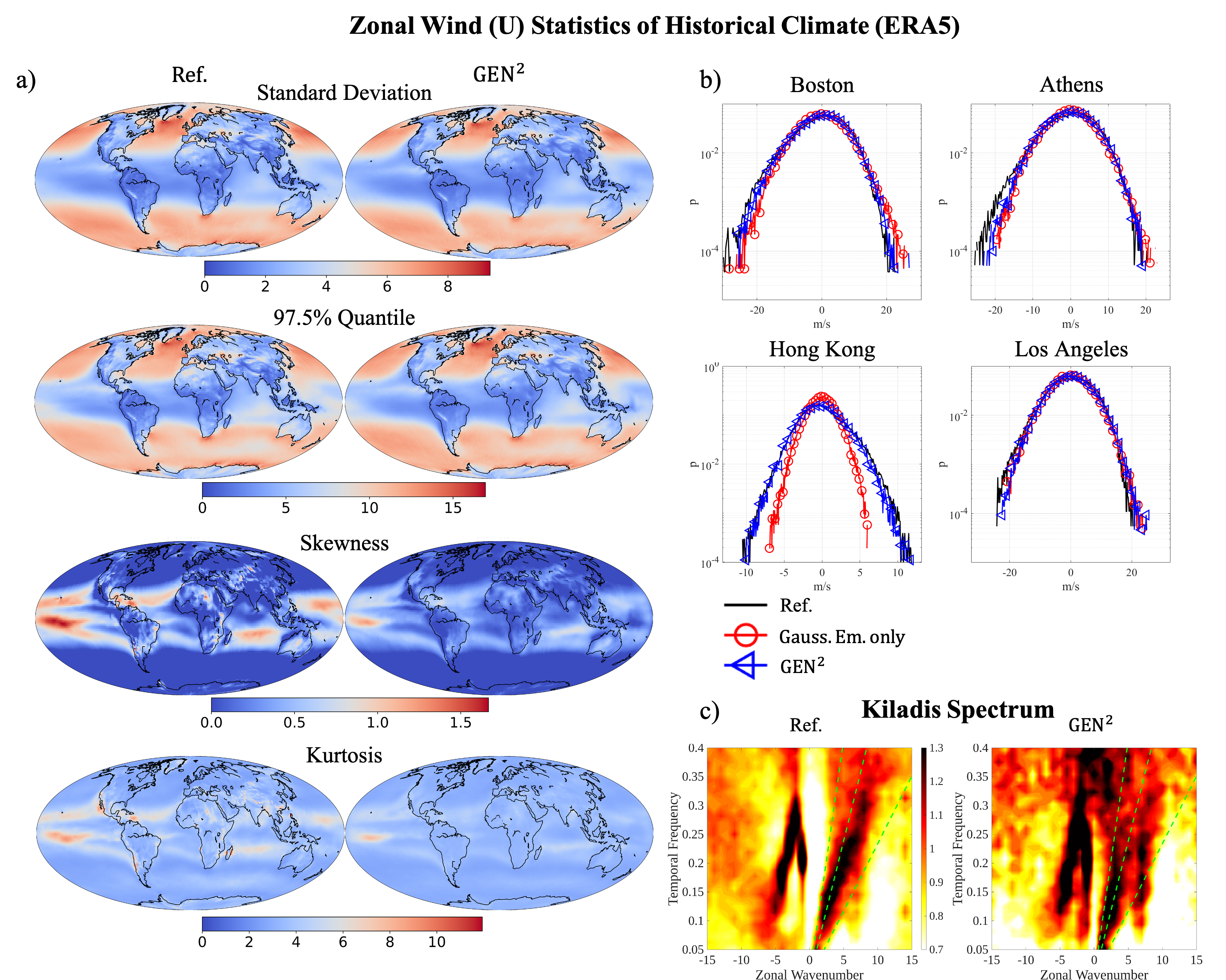}
    \caption{Zonal wind statistics of historical climate, evaluated using 1979-2018 reference ERA5 data and GEN$^2$ prediction.
    (a) Local standard deviation, 97.5$\%$ quantile, skewness and kurtosis.
    (b) Log probability density functions at areas including Boston, Athens, Hong Kong, and Los Angeles. Black: reference; Red: Gaussian emulator only; Blue: GEN$^2$.
    (c) Wheeler-Kiladis space time spectra. Green dashed lines: Kelvin waves with depth $H = 12m, 50m, 150m$ (Phase speed of Kelvin waves is $\sqrt{gH}$).
    }
    \label{fig:era5_1}
\end{figure}

We first validate our proposed modeling framework on historical ERA5 reanalysis data \citep{hersbach2020era5}.
The traininig dataset includes $U,V,T,Q$ from 1979 to 2018.
The temporal sampling frequency is three hours, and the data are projected onto a $1.5^{\circ} \times 1.5^{\circ}$ resolution grid, i.e. approximately $100$km.
Once trained, the GEN$^2$ model is used to generate a 1979-2018 trajectory, and the climate of this predicted trajectory is compared against the actual ERA5 data.
In figure \ref{fig:era5_1}, we show several metrics illustrating the ability of our model to capture the full richness of the ERA5 data.
Unless otherwise stated, these metrics are computed using the fluctuation fields, defined as the deviation from the known climatological mean.
All the metrics are temporally averaged from 1979 to 2018.
More detailed definitions are provided in Supplementary Methods (section 3).
Figure \ref{fig:era5_1}(a) compares the 40-year standard deviation, $97.5\%$ quantile, skewness, and kurtosis of the zonal wind ($U$).
In all cases, our model accurately predicts both the  qualitative and quantitative structure of the statistics -- although the skewness is slightly underestimated in the Pacific Ocean.
To systematically compare the full statistics, especially the extreme events, in panel \ref{fig:era5_1}(b), we plot the probability density function (PDF) of $U$ in log scale at four representative locations - Los Angeles, Boston, Athens, and Hong Kong.
The prediction of the conditional Gaussian emulator, without ML correction, is also provided for reference.
Interestingly, the conditional Gaussian emulator itself is already capable of capturing the distributions relatively well at locations where the distributions are weakly non-Gaussian.
The ML debiasing step maintains or slightly improves the prediction at these locations.
At Hong Kong, where the PDF is more strongly skewed, the ML debiasing step significantly corrects the tails of the distributions. More examples illustrating the debiasing power of the ML correction are included in Supplemental Information (Supplementary figure 3, 4 and table 1-4).
Beyond these single-point statistics, we also evaluate the spatio-temporal coherence of the predicted fields, by plotting the Wheeler-Kiladis spectrum \citep{wheeler_convectively_1999} of $U$, which quantifies the dispersion relationships of equatorial waves. As shown in figure \ref{fig:era5_1}(c), our model captures the characteristic frequency-wavenumber correlations corresponding to Kelvin waves observed in the data \citep{wheeler_convectively_1999,kiladis_convectively_2009} -- a remarkable observation given that our model includes no physics to enforce these dispersion relations.

\begin{figure}[h]
    \centering
    \includegraphics[width=0.95\textwidth]{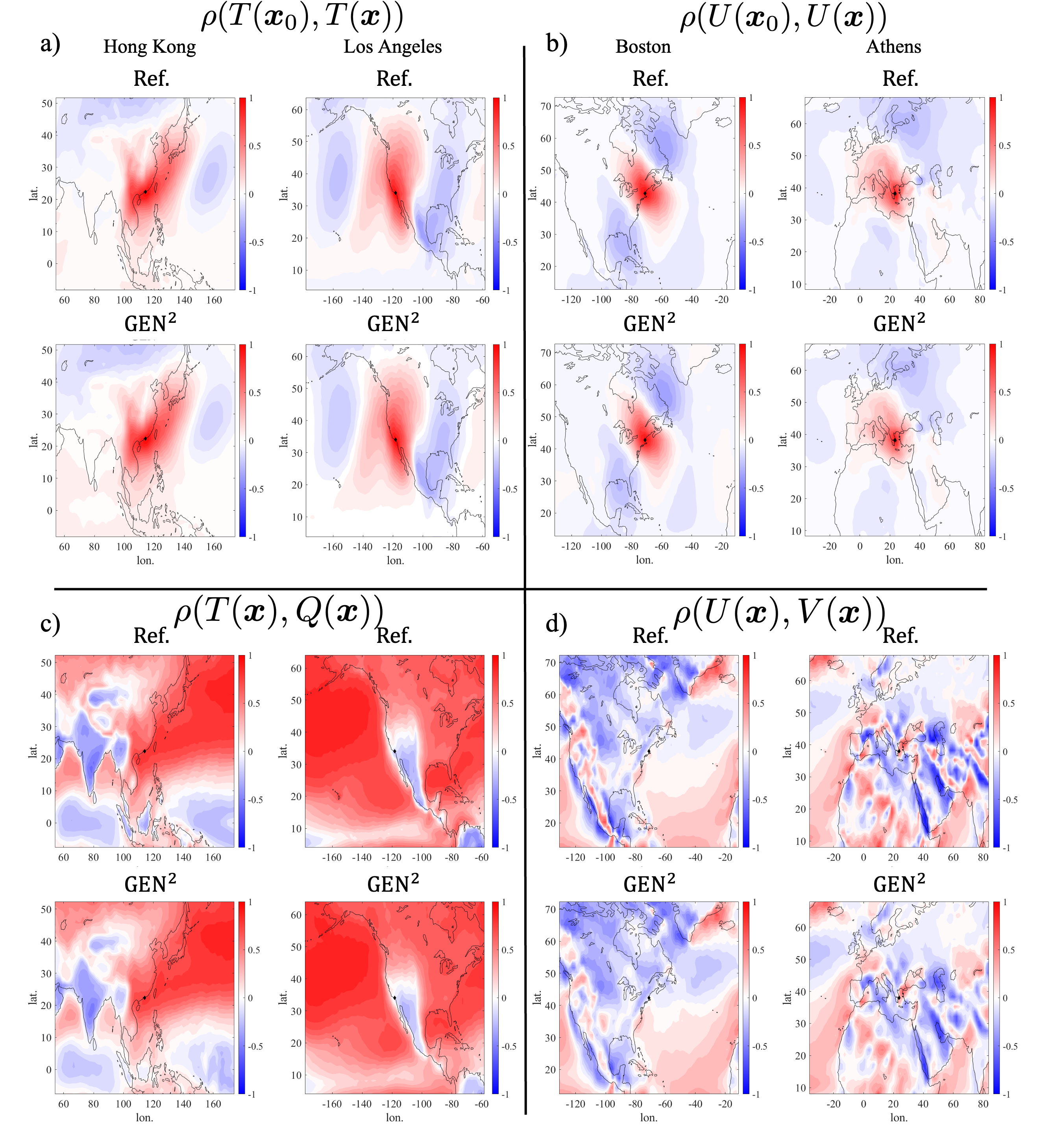}
    \caption{Regional correlation coefficients. (a,b) Two point correlations of temperature ($T$) and zonal wind ($U$), centered at four different cities. (c,d) Local cross-variable correlations between (c) temperature and humidity, and (d) zonal and meridional wind speed. For each sub-figure reference ERA5 data is shown in the top row and the GEN$^2$ prediction in the bottom row. All results are evaluated over the 40 year period 1987-2018 and at surface elevation.}
    \label{fig:era5_2}
\end{figure}

To further quantify the structure of the predicted fields, we compute the spatial two point correlation coefficients of temperature $\rho(T(\boldsymbol{x}_0),T(\boldsymbol{x}))$ and zonal wind $\rho(U(\boldsymbol{x}_0),U(\boldsymbol{x}))$, centered at each of the four previously analyzed locations. The results are shown in figure \ref{fig:era5_2} (a,b).
Moreoever, the cross-variable correlations at the same location, including zonal-meridonal wind correlation $\rho(U(\boldsymbol{x}),V(\boldsymbol{x}))$ as well as temperature-humidity correlation $\rho(T(\boldsymbol{x}),Q(\boldsymbol{x}))$, are plotted in figure \ref{fig:era5_2} (c,d).
Accurately capturing these correlations is crucial for accurately quantifying the risks of extreme weather events, which often occur due the concurrent incidence of extreme excursions in multiple climate variables, such as droughts being characterized by high temperatures and low humidity \citep{bevacqua_advancing_2023,zscheischler_future_2018,raymond_understanding_2020,robinson_increasing_2021}.
Again, our model captures the structure of the underlying data exceptionally well, see for example the negative correlation between temperature and humidity in India and the southwestern United States -- both places known to be susceptible to drought and extreme heat. We also capture the highly nontrivial wind patterns quantified by the correlation of $U$ and $V$ over both Europe and the United States.

\subsection{Climate Change Scenarios}
Having now demonstrated the capability and flexibility of our approach to reproduce the highly nontrivial structure of the historical climate, we now apply our method to forward looking climate change scenarios. Specifically, we consider the MPI-ESM1-2-LR model outputs of the Coupled Model Intercomparison Project Phase 6 (CMIP6) as our reference data.
Such a choice is based on two considerations. First, the MPI model dataset has multiple ensemble members and climate change scenarios available. Second, this model has demonstrated adequate skill in the quantification of climate extremes in a recent benchmarking study of CMIP6 models \citep{Wehner2020characterization}. 
We focus on four climate change scenarios, SSP126, SSP245, SSP370, and SSP585, each corresponding to a different level of global emissions, and accordingly a different trajectory of global mean temperature.

The crucial test of any data-driven model is its ability to extrapolate beyond the distribution of the data seen in training. We demonstrate this capability in two ways, first we will show the ability to extrapolate statistics \textit{within} a single climate change scenario, and second the ability to extrapolate to unseen scenarios. We train on 1 realization of the most extreme emission scenario, SSP585, and evaluate our model on 10 realizations of the climate under all 4 warming scenarios -- both the SSP585 scenario seen in training and the three other unseen scenarios.

\begin{figure}
    \centering
    \includegraphics[width=0.95\textwidth]{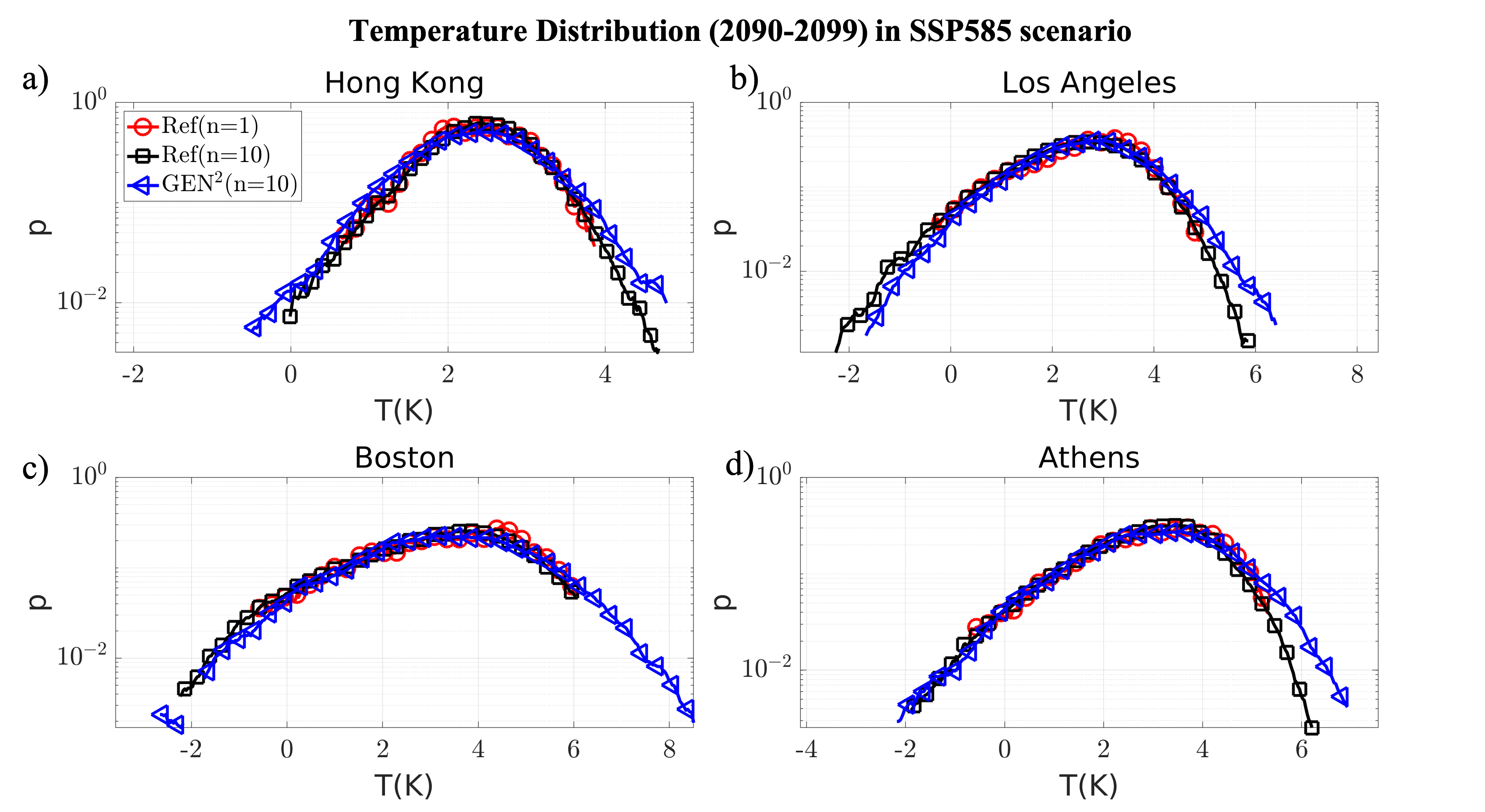}
    \caption{Demonstration of extrapolation within warming scenario. Probability densities of temperature in (a) Hong Kong , (b) Los Angeles, (c) Boston, and (d) Athens in 2090-2099 under SSP585 warming scenario. The ensemble statistics of 10 realizations of the reference data and GEN$^2$ prediction are shown in black squares and blue triangles respectively. The density of the single realization of reference data used to train the GEN$^2$ is shown in red circles. Results labeled ``Ref.'' represent reference simulation data, ``GEN$^2$'' represent our model predictions.}
    \label{fig:cmip6_0}
\end{figure}

We first demonstrate in figure \ref{fig:cmip6_0} the ability of our model to extrapolate within scenario. Here we show the probability densities of temperature fluctuations in Hong Kong, Los Angeles,  Boston, and Athens in 2090-2099 under the SSP585 warming scenario. The PDFs are computed by Monte Carlo sampling and smoothed by a moving average filter to improve readability.
Although only one member is used for training (red circles), the generated 10 realizations (blue triangles) successfully capture the tail statistics of the true 10 members (black squares).
If the underlying system is ergodic, this type of extrapolation from 1 to multiple realizations is equivalent to training the model on a short time window and testing on a longer time window, as demonstrated in \citet{barthel_sorensen_non-intrusive_2024,barthel_sorensen_probabilistic_2024}.

\begin{figure}
    \centering
    \includegraphics[width=0.95\textwidth]{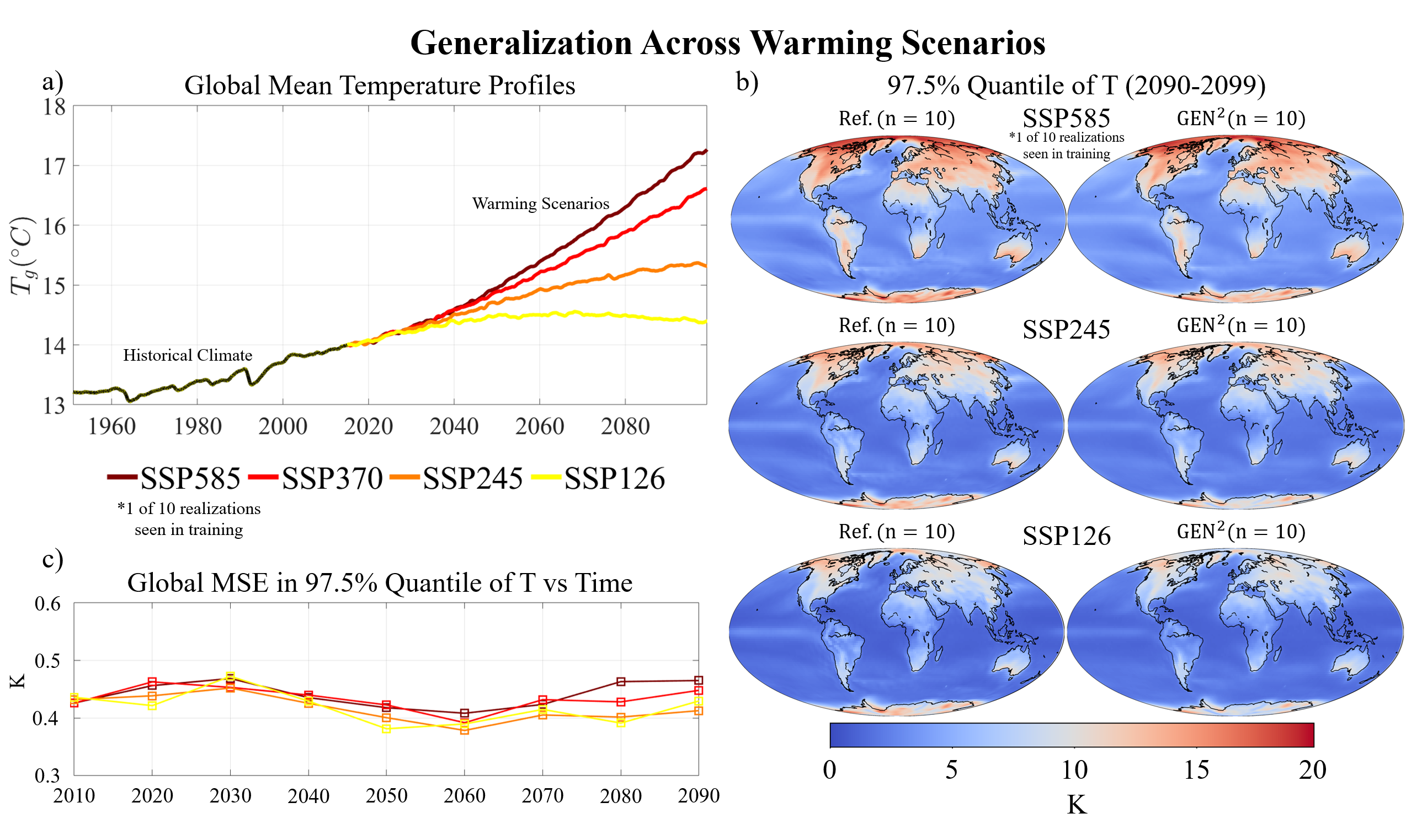}
    \caption{Demonstration of generalizability across climate change scenarios. (a) Illustration of Global mean temperature profile (which serve as inputs to our model) corresponding to various climate change scenarios - figure shows ensemble average over 10 realizations. Only one of the realization of the SSP585 scenario is used for training. (b) Global field of 97.5$\%$ quantiles of  10 realizations of temperature fluctuations for 3 different warming scenarios for the years 2090-2099. Results labeled ``Ref.'' represent reference simulation data, ``GEN$^2$'' represent our model predictions. (c) Global root-mean-square error of the same temperature fluctuation quantiles over time. Each data point represents the RMSE of the two quantile fields in (b) computed for each decade. }
    \label{fig:cmip6_1}
\end{figure}

We next demonstrate the ability of our approach to extrapolate beyond the scenario seen in training, as shown in figure \ref{fig:cmip6_1}. Figure \ref{fig:cmip6_1}(a) illustrates the evolution of global mean temperature corresponding to the four emissions scenarios studied here. Each curve shows the ensemble average over 10 members. Figure \ref{fig:cmip6_1}(b) compares the $97.5\%$ quantile of temperature predicted by our model to the reference data at the end of the century (2090-2099) for three different scenarios SSP125, 245, 585. Since these are the quantiles of the climatological-mean-subtracted fields, the peak values are observed at the poles (as opposed to the equator), indicating that the impacts of extreme temperature fluctuations will be most pronounced in the polar regions under strong global warming. Figure \ref{fig:cmip6_1}(c) shows the root-mean square error (RMSE) in the predicted quantiles for each scenario as a function of time -- that is to say the RMSE of the fields shown in figure \ref{fig:cmip6_1}(b) computed for each decade. Our model is able to successfully and consistently predict the quantiles in unseen scenarios, achieving comparably low error  
$(< 0.5K)$ across all scenario despite the fact that only data from one of the scenarios was seen in training. This is a critical ability in any climate emulator, as it means that new scenarios of interest can be reliably investigated without additional numerical simulations or retraining of exiting data-driven models.

\begin{figure}
    \centering
    \includegraphics[width=0.95\textwidth]{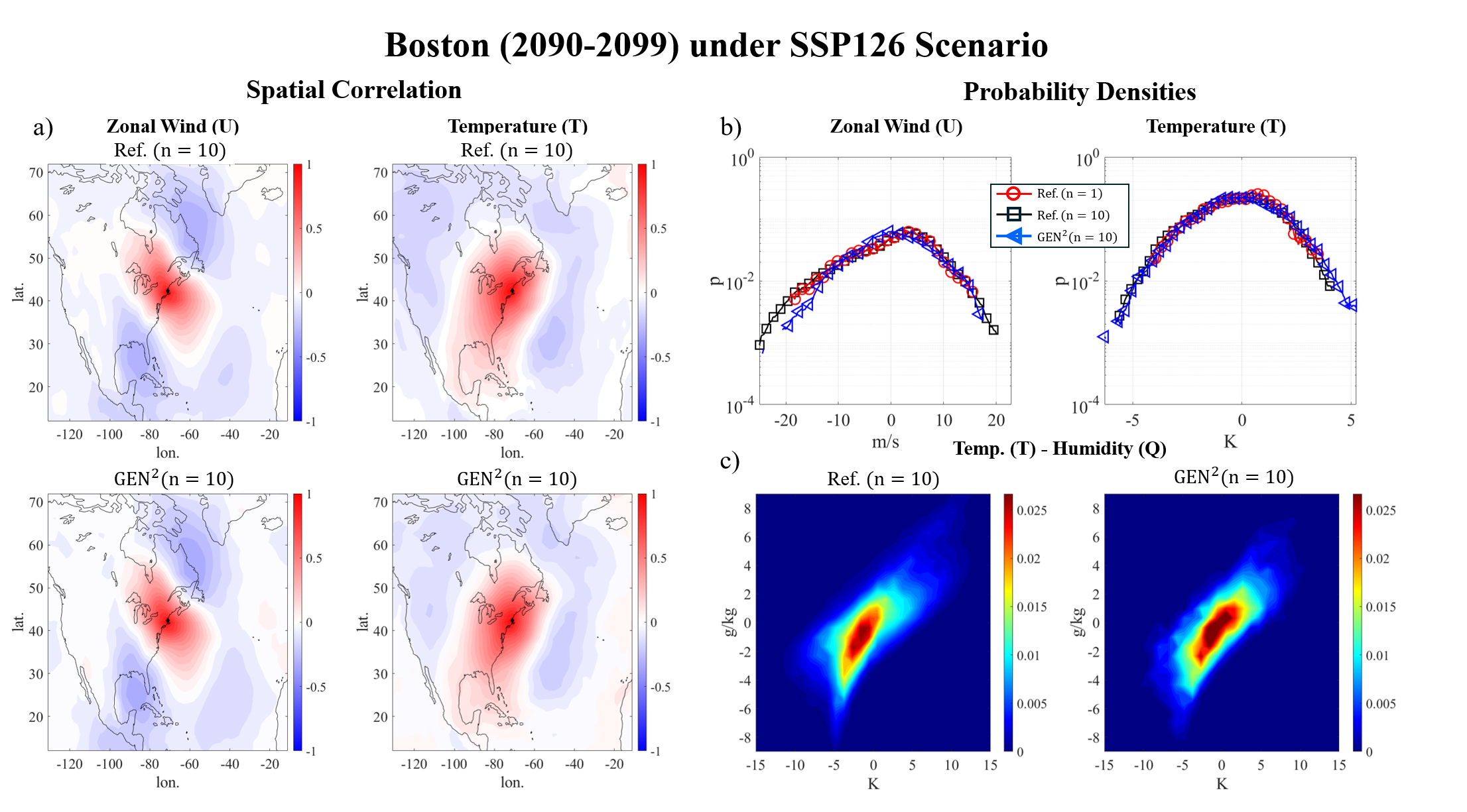}
    \caption{Demonstration of regional climate statistics. (a) Spatial correlation coefficient, (b) probability density and (c) joint probability density of zonal wind (U) and temperature (T) over Boston. Results labeled ``Ref.'' represent reference simulation data, ``GEN$^2$'' represent our model predictions.  All statistics are computed from 10 ensemble members for the years 2090-2099. }
    \label{fig:cmip6_2}
\end{figure}

To further highlight the ability of our model to replicate warming scenarios not seen in training, we zoom in to the region centered around Boston under the SSP126 scenario, which is the most dissimilar from the SSP585 data seen in training.
Figure \ref{fig:cmip6_2} shows the two point correlations (panel a), PDFs of $U$ and $T$ (panel b), and the joint PDF of $T$ and $Q$ (panel c).
All these statistics are computed using the fluctuation fields and averaged from 2090 to 2099.
In all cases our model prediction captures the highly non-Gaussian and non-isotropic structure observed in the reference data.
Our model again manages to extrapolate the tails in the local PDFs (panel b).
The shape of joint PDFs in (c) indicate that fluctuations in humidity are positively correlated with fluctuations in temperature, representing the relative prevalence of dry cold snaps and humid heatwaves -- patterns not unfamiliar to residents of New England and accurately predicted by GEN$^2$ approach.


\FloatBarrier
\section{Discussion}\label{sec:disc}
The prediction-correction framework presented here provides a more accessible and computationally affordable alternative to Earth System Models for generating large ensembles of climate realizations.
Once trained, our GEN$^2$ model generates predictions at a cost of approximately 720 simulated years-per-day (YPD), whereas Earth System Models achieve approximately 40 YPD \cite{golaz2022doe}. Compared to traditional data-driven emulators such as pattern scaling, our model is able to capture  non-Gaussian statistics as well as significantly improve the estimation of higher-order moments, spatial correlations, and highly non-trivial spatio-temporal features such as the Kiladis spectrum.
Compared with other ML-based emulators that may offer sufficient flexibility to capture higher-order features, our approach achieves a lower computational cost of both training and inference, by obtaining a first-pass prediction using the conditional Gaussian emulator and limiting the more costly ML step to a debiasing operation. Another benefit of the initial Gaussian emulation is the ability to incorporate different physics by customizing which statistics are regressed on. For example, while we chose to enforce temporal coherence over several days, with sufficient data one could also choose to enforce inter-annual or season-to-season correlations.
This would generally be much more difficult with purely ML based emulators.

The GEN$^2$ framework can be generalized along different pathways to further improve its accuracy, and we outline two possibilities below.
First, our debiasing operator is based on an image diffusion model, which operates on each snapshot in time independently and thus can not correct biases in temporal dynamics.
Debiasing in time would require the use of a video diffusion model, but this approach would significantly increase the training and inference cost.
Second, quantifying very localized weather statistics requires additional localized super-resolution (or downscaling) of the predicted fields -- a feature not incorporated in our framework. However, many existing techniques \cite{wan_debias_2023} could be directly applied to the output of our model, and this remains a topic of ongoing work.
In summary, the GEN$^2$ framework is capable of reconciling the computational cost constraints with the need for accurate climate extreme emulation.
This framework serves as a guide for future developments of climate emulators, enabling policymakers and stakeholders to explore the parameter space of emission trajectories and interventions.

\FloatBarrier
\section{Methods}\label{sec:methods}
Consider the state variable of the climate system as a function of space and time, $\boldsymbol{u}(\boldsymbol{x},t)$. Its underlying dynamical system is generally chaotic and high-dimensional, which makes it intractable to capture the full dynamics using deterministic data-driven models.
Faced with this challenge, we seek a stochastic model that approximates the dynamics as a function of the emission scenario, quantified by the global mean temperature $T_g$.
Specifically, we seek to parameterize the following conditional probability distribution
\begin{equation}\label{eq:model_form}
    p\left(\boldsymbol{u}(\boldsymbol{x},t)|T_g(t)\right) \sim\mathcal{G}_{\theta}[T_g(t)]
\end{equation}
through a prediction-correction process
\begin{align}
\hat{\boldsymbol{q}}(\boldsymbol{x},t) & \sim \mathcal{G}_{\theta,1}\left[T_g(t)\right] \\
    \hat{\boldsymbol{u}}(\boldsymbol{x},t)& \sim\mathcal{G}_{\theta,2}\left[\hat{\boldsymbol{q}}(\boldsymbol{x},t)\right],
\end{align}
where $\mathcal{G}_{\theta,1}$ is a conditional Gaussian emulator, linearly driven by $T_g$, and  $\mathcal{G}_{\theta,2}$ is a general nonlinear non-Gaussian stochastic model, parametrized by $\theta$. 
In general we wish the model (\ref{eq:model_form}) to fulfill three main aims. (1) Forward evaluation of the model must be cheap so that a large number of ensembles can be rapidly generated. (2) The model must be stable over arbitrarily long time horizons. (3) The model must be capable of extrapolating beyond the distribution of the data seen in training. A diagram describing the full model is shown in figure \ref{fig:method_diagram}.

\subsection{Step 1: Conditional Gaussian Emulation}
\label{subsec:emulator}
The conditional Gaussian emulator is built on the framework introduced by \citet{wang2023emulator}.
We extend this approach to emulate multiple variables and capture the correct spatio-temporal correlations.
The emulated state, $\hat{\boldsymbol{q}}(\boldsymbol{x},t)$, is constructed as the superposition of the climatological mean $\bar{\boldsymbol{u}}$ and the PCA modes,
\begin{equation}
    \label{eq:q_pca}
    \hat{q}_k(\boldsymbol{x},t) = \bar{u}_k(\boldsymbol{x},t) +  \sigma_{g,k} \sum_{i=1}^I \hat{a}_i(t) \phi_k^{(i)}(\boldsymbol{x}).
\end{equation}
where the subscript $k = 1,2,3,4$ corresponds to $U, V, T, Q$ components, $\sigma_{g,k}$ is the globally-averaged standard deviation, and $\phi_k^{(i)}$ is the $i^{\text{th}}$ PCA mode. The quantities $\bar{\boldsymbol{u}}$, $\sigma_{g,k}$, and $\phi_k^{(i)}$ are all assumed to be known.

The time series of PCA coefficients are modelled as Gaussian processes conditioned on $T_g$, which characterizes the climate change.
Specifically, the modelled time series, 
\begin{equation}
	\label{eq:model_a}
\hat{a}_{i}(t) = \hat{\mu}_{i}\left(T_{g} \right) + \hat{\sigma}_{i}\left(T_{g} \right) \hat{\eta}_{i}(t),
\end{equation}
consist of the seasonal mean $\hat{\mu}_{i}\left(T_{g} \right)$ and daily fluctuations $\hat{\eta}_{i}(t)$ scaled by the seasonal standard deviation $ \hat{\sigma}_{i}\left(T_{g} \right)$.
Here $\hat{\mu}_{i}$ and $\hat{\sigma}_{i}$ are assumed piecewise constant in each season and varying linearly with the seasonal average of the global mean temperature $T_{g}$ (as demonstrated in Supplementary Figure 1). The parameters of the linear models are estimated by performing least square regression on the data in each season respectively.
The daily fluctuations $\boldsymbol{\eta}(t)$ are modelled as zero-mean multivariate Gaussian processes, whose covariance matrces are assumed constant in each season and estimated from data.
Once trained, this emulator takes as input a scalar valued time series of global mean temperature $T_{g}(t)$ and outputs a prediction of the full spatio-temporal evolution $\hat{\boldsymbol{q}}(\boldsymbol{x},t)$. A more detailed mathematical description of the model is included in the Supplementary Methods.

In order to accurately represent the global climate system while maintaining computational efficiency, we only keep the first 500 PCA modes that account for approximately 80\% of the total variance of the training data.
As such, the bias of the emulator arises from two sources: the non-Gaussian statistics of the first 500 modes and the ignored higher-order modes. 
The latter generally correspond to the small-scale and more extreme events. These biases will be corrected through the deep learning model in the next section.

\subsection{Step 2: Nudged Emulation and Non-Gaussian Correction}\label{subsec:ml}
Due to the chaotic nature of the Earth system and herein the stochasticity of the emulator, the emulated state $\hat{\boldsymbol{q}}(\boldsymbol{x},t)$ will significantly deviate from contemporaneous reference $\boldsymbol{u}(\boldsymbol{x},t)$, making it fundamentally difficult to learn a map $\mathcal{G}_{\theta,2}$ from $\hat{\boldsymbol{q}}(\boldsymbol{x},t)$ to $\boldsymbol{u}(\boldsymbol{x},t)$.
To address this challenge, we construct a \emph{nudged} trajectory $\hat{\boldsymbol{q}}^{\nu}$ that stays close to the reference trajectory $\boldsymbol{u}$ while approximately satisfying the equations of the conditional Gaussian emulator.
If we re-formulate the emulator (\ref{eq:model_a}) as a dynamical system,
\begin{equation}
    \label{eq:dadt}
    \frac{\mathrm{d}\hat{\boldsymbol{a}}}{\mathrm{d}t} = f(\hat{\boldsymbol{a}},t),
\end{equation}
the nudged PCA time series are defined as,
\begin{equation}
    \label{eq:dadt_nudge}
    \frac{\mathrm{d}\hat{\boldsymbol{a}}^{\nu}}{\mathrm{d}t} = f(\hat{\boldsymbol{a}}^{\nu},t) - \frac{1}{\tau}\left(\hat{\boldsymbol{a}}^{\nu} - \boldsymbol{a}  \right).
\end{equation}
The vector $\boldsymbol{a}$ denotes the first 500 modes that are included into the emulator.
Compared with equation (\ref{eq:dadt}), the nudged emulator (\ref{eq:dadt_nudge}) features an additional feedback term that forces $\hat{\boldsymbol{a}}^{\nu}$ towards the true PCA coefficients $\boldsymbol{a}$.
From a physical perspective, the nudging term in (\ref{eq:dadt_nudge}) enforces the slow dynamics of $\hat{\boldsymbol{a}}^{\nu}$ to follow the reference, while allowing fast and more extreme dynamics to freely evolve \cite{barthel_sorensen_non-intrusive_2024}.
The nudging timescale $\tau$ is a user-defined parameter. Generally, $\tau$ should be selected to ensure that the nudging term is an order of magnitude than the other terms in the governing equations. 
In this study, $\tau$ is set as six hours for both ERA5 and CMIP6 data, which is also consistent with previous studies \cite{barthel_sorensen_non-intrusive_2024}.
Combining $\hat{\boldsymbol{a}}^{\nu}$ with the PCA mode shapes gives us the nudged spatial-temporal fields $\hat{\boldsymbol{q}}^{\nu}(\boldsymbol{x},t)$, which will be utilized together with the reference data $\boldsymbol{u}(\boldsymbol{x},t)$ to learn $\mathcal{G}_{\theta,2}$.

The debiasing operator $\mathcal{G}_{\theta,2}$ is parameterized as a conditional score-based diffusion model \cite{song_score-based_2021,bischoff_unpaired_2023}.
This probabilistic approach accounts for the potentially non-unique relationship between $\hat{\boldsymbol{q}}^{\nu}$ and $\boldsymbol{u}$.
Once the diffusion model is trained to learn $p(\boldsymbol{u} | \hat{\boldsymbol{q}}^{\nu})$ , it can take any free-running emulation $\hat{\boldsymbol{q}}(\boldsymbol{x},t)$ as an input or conditional information to produce the debiased fields $\hat{\boldsymbol{u}}(\boldsymbol{x},t)$.
For example, in CMIP6 data, $\hat{\boldsymbol{q}}(\boldsymbol{x},t)$ could come from a realization of the climate change scenario that is unseen during training, in order to evaluate the capability of our framework to extrapolate beyond the training scenarios. More details about the conditional diffusion model are provided in the Supplemental Methods.
\backmatter

\bmhead{Supplementary information}

This article has accompanying supplemental materials.



\bmhead{Acknowledgments}

This research was partially supported by the Vannevar Bush grant N000142512059, as well as the project Bringing Computation to the Climate Challenge (BC3), supported by Schmidt Sciences through the MIT Climate Grand Challenges.












\begin{appendices}
\section{Stochastic Emulator}\label{app:emulator}
Here we describe in detail the first part of our climate modeling framework, the linear stochastic emulation. In summary, the emulator takes as input a time series of the global mean temperature $T_g(t)$ and outputs a time series of the local state of the climate $\boldsymbol{u}(\boldsymbol{x},t) = [u_1,u_2,u_3,u_4]^{\top} = [U(\boldsymbol{x},t), V(\boldsymbol{x},t), T(\boldsymbol{x},t), Q(\boldsymbol{x},t)]^{\top}$ where $U,V,T, Q$ are the zonal and meridional wind speeds, temperature and humidity respectively and the spatial dimensions $\boldsymbol{x} = (\theta,\varphi)$ are the longitude and latitude, $\theta \in [-\pi/2,\pi,2]$ and $\varphi \in [0,2\pi)$.
The time step size of $t$ is three hours for ERA5 dataset and one day for the CMIP6 MPI model.
Consistent with the formulation of modern climate models -- and to reduce the data to a manageable size -- our model operates at a fixed altitude, and thus the spatial dimension is 2D. We focus here exclusively on the near-surface climate, but our model could be directly applied to any altitude.

Stated succinctly, our approach consider a principal component analysis (PCA) of the climate data $\boldsymbol{u} (\boldsymbol{x},t)= \sum_j a_j(t)\boldsymbol{\phi}_j(\boldsymbol{x})$ and  attempts to model the temporal coefficients $a_j(t)$ for a given spatial basis $\boldsymbol{\phi}_j(\boldsymbol{x})$. Our emulator is therefore built on three fundamental assumptions: 
\begin{enumerate}
    \item The PCA basis $\boldsymbol{\phi}_j(\boldsymbol{x})$ computed from the climate during a sufficiently long time period (e.g. historical and SSP5-8.5 scenario) remains an efficient basis for describing other future climate change scenarios;
    \item The seasonal mean and variance of the coefficients $a_j(t)$ vary linearly with global mean temperature. 
    \item The statistics of daily fluctuations, given the season, are independent of the year and the climate change scenarios.
\end{enumerate}
The construction of the emulator can be divided into two distinct steps: dimensionality reduction and stochastic modeling of PCA time series. The emulator is then nudged towards the observation data to facilitate machine-learning-based debiasing. We now describe each of these steps in detail.

\begin{figure}
    \centering
    \includegraphics[width=\textwidth]{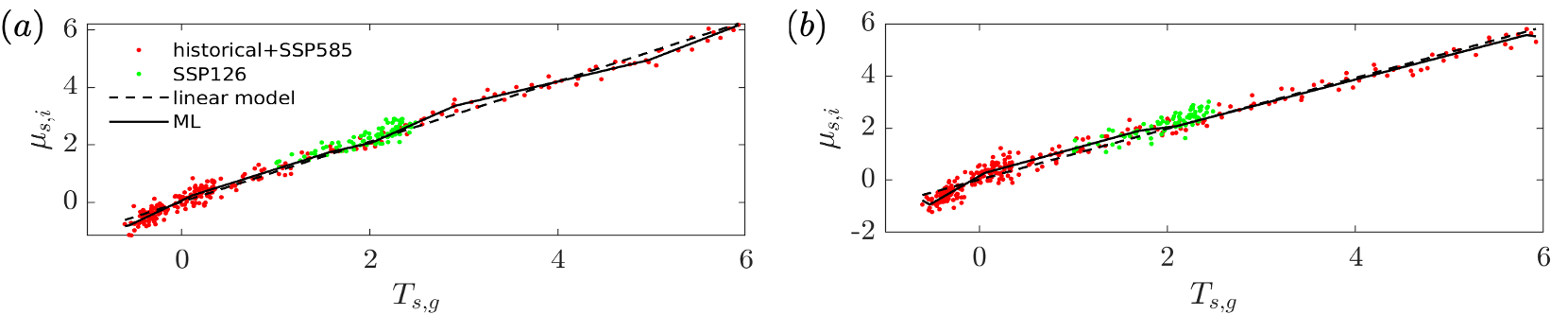}
    \caption{Jun-Aug mean of ($a$) the first and ($b$) second PCA coefficients in each year of CNRM-CM6-1-HR dataset, from 1850 to 2100, plotted versus the global mean temperature. Red dots: true seasonal mean obtained from the historical and SSP5-8.5 scenario. Green dots: SSP1-2.6 scenario. Black dashed line: linear regression; Solid line: machine-learned function.}
    \label{fig:mu_Tg}
\end{figure}

\subsection{Dimensionality Reduction}

First we describe how the spatial PCA basis $\boldsymbol{\phi}_j(\boldsymbol{x})$, which provides the structure for our emulator, is computed.
Given a dataset consisting of $N$ years, we extract the climatological mean $\bar{\boldsymbol{u}}(\boldsymbol{x},t)$,  defined as phase-average of $\boldsymbol{u}$ on the same calendar day (e.g. Jan 1st),
\begin{equation}
    \label{eq:q_mean}
    \bar{\boldsymbol{u}}(\boldsymbol{x},t) = \frac{1}{N}\sum_{n=0}^{N-1} \boldsymbol{u}(\boldsymbol{x},t + nT), \quad 1 \leq t \leq T.
\end{equation},
where the period $T$ is one year.
When the emulator is trained on the daily maximum data from the MPI model, the climatological mean (\ref{eq:q_mean}) only quantifies the seasonal cycle.
For the three-hourly ERA5 data, $\bar{\boldsymbol{u}}$ accounts for not only the seasonal variation but also the diurnal cycles.
To obtain the scaling of each state variable, we compute the global-and-time-averaged standard deviation,
\begin{equation}
    \label{eq:sigma_g}
    \sigma_{g,k} = \left[ \frac{1}{\mathcal{T}S}\int_0^{\mathcal{T}} \int_S \left(u_k(\boldsymbol{x},t) - \bar{u}_k (\boldsymbol{x},t)\right)^2 \cos\theta d\theta d\varphi dt \right]^{1/2}.
\end{equation}
The notations $\mathcal{T}$ and $S$ are the duration of training window and the Earth's surface, respectively.
The data are then centered to have zero climatological mean and scaled by the global standard deviation,
\begin{equation}
    \label{eq:q_fluct}
    u_k^{\prime}(\boldsymbol{x},t) = \left(u_k(\boldsymbol{x},t) - \bar{u}_k(\boldsymbol{x},t)\right) / \sigma_{g,k}.
\end{equation}

Now that each component of $q^{\prime}_k$ has the same order of magnitude, we construct its spatial covariance function,
\begin{equation}
    \label{eq:spatial_cov}
    \mathcal{R}_{jk}(\boldsymbol{x},\boldsymbol{x}^*) = \frac{1}{\mathcal{T}} \int_0^T u^{\prime}_j(\boldsymbol{x},t) u^{\prime}_k(\boldsymbol{x}^*,t) dt, \quad j,k = 1,2,3,4.
\end{equation}
The PCA modes are acquired by solving the eigenvalue problem,
\begin{equation}
    \label{eq:EOF}
    \int_{S} \sum_{k}\mathcal{R}_{jk}(\boldsymbol{x},\boldsymbol{x}^*) \phi_k (\boldsymbol{x^*}) \cos{\theta} d\theta d\varphi = \lambda \phi_j(\boldsymbol{x}), \quad j = 1,2,3,4,
\end{equation}
This set of equations has multiple solutions $(\lambda^{(i)},\boldsymbol{\phi}^{(i)})$, $i = 1,2,3,\ldots$, which are the PCA eigenvalues and mode shapes, respectively.
Without loss of generality we rank the eigenpairs such that the eigenvalues, which represent variance, satisfy $\lambda_1>\lambda_2 >\ldots > \lambda_I$.
The temporal PCA coefficients which govern the time dependence of the spatial PCA modes are found by projecting the  
normalized fluctuation field onto $\boldsymbol{\phi}^{(i)}$,
\begin{equation}
    \label{eq:EOF_proj}
    a_i(t) = \int_{S} \sum_k u^{\prime}_k(\boldsymbol{x},t) \phi^{(i)}_k(\boldsymbol{x}) \cos{\theta} d\theta d\varphi.
\end{equation}
The state of the climate can then be expressed as superposition of PCA modes,
\begin{equation}
    u_k(\boldsymbol{x},t) = \bar{u}_k(\boldsymbol{x},t) + \sigma_{g,k} \sum_{i=1}^{I} a_i(t) \phi_k^{(i)}(\boldsymbol{x}).
\end{equation}

When the number of PCA modes $I$ is equal to the number of grid points or the number of snapshots, whichever is smaller, we recover the full field, and any smaller value of $I$ represents a truncation.
In this work, we always retain 500 PCA modes, which represent 79.6\% of the total variance of 1979-2018 ERA5 data and 78.2\% of 1950-2100 MPI data.
To reiterate, we assume that the climatological mean $\bar{u}_k(\boldsymbol{x},t)$, global standard deviation $\sigma_{g,k}$, and PCA mode shapes $\phi_k^{(i)}$ are unchanged with time or future scenarios. As a result, we focus purely on modeling $a_i(t)$, and the emulated state is written as,
\begin{equation}
    \label{eq:qhat}
    \hat{q}_k(\boldsymbol{x},t) = \bar{u}_k(\boldsymbol{x},t) + \sigma_{g,k} \sum_{i=1}^{500} \hat{a}_i(t) \phi_k^{(i)}(\boldsymbol{x}).
\end{equation}
Here the notations with $\hat{\cdot}$ are emulated quantities. 

\subsection{Stochastic emulator of PCA time series}
\label{sec:pca_time_series}

\subsubsection{Seasonal Decomposition}
Our goal is to construct a time series of $\hat{a}_i(t)$ that \textit{statistically} resembles the reference data $a_i(t)$. Although we have removed the climatological mean, the statistics of $a_i(t)$ still exhibit seasonal variation that is important to take into account. Therefore, we divide $a_i(t)$ into four seasons $a_{s,i}(t)$  -- of approximately equal length -- and model them separately where the additional subscript $s = 1,2,3,4$ represents winter (Dec-Feb), spring (Mar-May), summer (Jun-Aug) and autumn (Sep-Nov). The number of days in each season is 90, 92, 92, and 91 respectively. 


\subsubsection{Formulation and Estimation of Model Parameters}
We postulate a decomposition of the time series of PCA coefficients,
\begin{equation}
    \label{eq:model_ahat}
    \hat{a}_{s,i}(t) = \hat{\mu}_{s,i}\left(T_{s,g} \right) + \hat{\sigma}_{s,i}\left(T_{s,g} \right) \hat{\eta}_{s,i}(t),
\end{equation}
which is a superposition of the seasonal mean $\hat{\mu}_{s,i}$ and fluctuations parameterized through an envelope of the seasonal variance $\hat{\sigma}_{s,i}^2$.
The seasonal mean and variance are assumed to be functions of the global mean temperature $T_{s,g}$, defined as the seasonal average of the daily $T_g$.
The time-dependent daily fluctuations in each season are modelled as autoregressive Gaussian processes $\hat{\eta}_{s,i}(t)$.
We will now discuss the formulation and computation of each of these terms in detail.

\noindent{\textbf{Linear Regression of Seasonal Mean and Variance.}}
For each season $s$ and each mode $i$, in the $n$th year, we compute the $T_{s,g}$ as well as the seasonal mean $\mu_{s,i}$ and variance $\sigma^2_{s,i}$ of the PCA coefficients of the reference data $a_{s,i}(t)$. Note that for each $s,i$, and $n$, the mean $\mu_{s,i}$ and variance $\sigma^2_{s,i}$ are constants -- we generally omit explicit notation of the year $n$ to avoid notational clutter. Grouping these values by season $s$ and mode $i$ allows us to perform a linear regression using $\{\mu_{s,i}(n),T_{s,g}(n)\}$ and  $\{\sigma^2_{s,i}(n),T_{s,g}(n)\}$
\begin{align} 
    \label{eq:lin_fit}
    \hat{\mu}_{s,i}(T_{s,g}) & = \hat{p}_{s,i,0} + \hat{p}_{s,i,1} T_{s,g} \nonumber \\
    \hat{\sigma}^2_{s,i}(T_{s,g}) & = \hat{q}_{s,i,0} + \hat{q}_{s,i,1} T_{s,g},
\end{align}
an assumption which is justified by the linear trends which have been observed in data by a number of sources \cite{tebaldi_pattern_2014,osborn_performance_2018} and illustrated in figure \ref{fig:mu_Tg}. 

\noindent{\textbf{Time Lagged Cross-Mode Covariance.}}
After extracting the linear trends of the seasonal mean and standard deviation in response to the global mean temperature, we remove these trends from the true PCA coefficients, resulting in the residuals 
$\eta_{s,i} = \left(a_{s,i}-\hat{\mu}_{s,i} \right) / \hat{\sigma}_{s,i}$.
To accurately capture the spatio-temporal dynamics, our model must reflect not only the contemporaneous correlations between different modes, e.g. $\eta_{s,1}(t)$ and $\eta_{s,2}(t)$, but also their correlations across time.
To this end, we define the \textit{time-lagged cross-mode covariance},
\begin{equation}
    \label{eq:Sigma}
    \mathbf{\Sigma}_s(m) = \frac{1}{\mathcal{T}_s} \int_{\mathcal{T}_s} \boldsymbol{\eta}_s(t) \boldsymbol{\eta}^{\top}_s(t+m \Delta t) dt, \quad m = 0, 1, \ldots M,
\end{equation}
where $\boldsymbol{\eta}_s = [\eta_{s,1}, \eta_{s,2}, \ldots, \eta_{s,m}]^{\top}$ is the vector of fluctuations of each PCA mode, $M \Delta t$ is maximum time lag considered, and $\mathcal{T}_s$ represents the set of time indices corresponding to season $s$ across all training years.

Now we want to model the observed fluctuations $\eta_{s,i}(t)$ as a multivariate Gaussian process $\hat{\eta}_{s,i}(t)$, which has the same covariance matrix $ \mathbf{\Sigma}_s(m)$ as $\eta_{s,i}(t)$.
To further simplify our notation, the subscript $s$ will be omitted.
Mathematically, we seek to construct an autoregressive model of order $M$,
\begin{equation}
    \label{eq:VAR}
    \hat{\boldsymbol{\eta}}(t) = \hat{\mathbf{\Psi}}_1 \hat{\boldsymbol{\eta}}(t-\Delta t) + \hat{\mathbf{\Psi}}_2 \hat{\boldsymbol{\eta}}(t-2\Delta t) + \cdots + \hat{\mathbf{\Psi}}_M \hat{\boldsymbol{\eta}}(t-M\Delta t) + \boldsymbol{\epsilon}(t)
\end{equation}
where the noise term is a multivariate Gaussian random vector $\boldsymbol{\epsilon} \sim \mathcal{N}(\mathbf{0}, \hat{\mathbf{R}})$.
The unknown matrices $\hat{\mathbf{\Psi}}_1, \hat{\mathbf{\Psi}}_2, \ldots, \hat{\mathbf{\Psi}}_M, \hat{\mathbf{R}}$ are solved such that the simulated process (\ref{eq:VAR}) satisfy the given covariance matrices with different time lags $\mathbf{\Sigma}(0),\mathbf{\Sigma}(1), \ldots \mathbf{\Sigma}(M \Delta t)$.
By multiplying both sides of (\ref{eq:VAR}) by $\hat{\boldsymbol{\eta}}(t-i\Delta t)$ and averaging in time, we can derive a set of equations, the so-called Yule-Walker equations,
\begin{equation}
    \label{eq:YuleWalker}
    \begin{bmatrix}
    \mathbf{\Sigma}(0) & \mathbf{\Sigma}^{\top}(1) & \cdots & \mathbf{\Sigma}^{\top}(M-1) \\
    \mathbf{\Sigma}(1) & \mathbf{\Sigma}(0) & \cdots & \mathbf{\Sigma}^{\top}(M-2) \\
    \vdots & \vdots & \ddots & \vdots \\
    \mathbf{\Sigma}(M-1) & \mathbf{\Sigma}(M-2) & \cdots & \mathbf{\Sigma}(0)
    \end{bmatrix}
    \begin{bmatrix}
    \hat{\mathbf{\Psi}}_1^{\top} \\ \hat{\mathbf{\Psi}}_2^{\top} \\ \vdots \\ \hat{\mathbf{\Psi}}_M^{\top}
    \end{bmatrix}
    =
    \begin{bmatrix}
    \mathbf{\Sigma}(1) \\ \mathbf{\Sigma}(2) \\ \vdots \\ \mathbf{\Sigma}(M)
    \end{bmatrix}.
\end{equation}
which may be readily solved for the $\hat{\mathbf{\Psi}}_j$ \citep{box_time_1994}. The corresponding noise covariance is then given by
\begin{equation}
    \label{eq:R_cov}
    \hat{\mathbf{R}} = \mathbf{\Sigma}(0) - \sum_{m=1}^M \hat{\mathbf{\Psi}}_m \mathbf{\Sigma}(m).
\end{equation}
After solving equations (\ref{eq:YuleWalker},\ref{eq:R_cov}), the matrices $\hat{\mathbf{\Psi}}_1, \hat{\mathbf{\Psi}}_2, \ldots, \hat{\mathbf{\Psi}}_M, \hat{\mathbf{R}}$ are substituted into the autoregressive model (\ref{eq:VAR}) to simulate the daily fluctuations. The complete procedures for running the emulator are summarized in Algorithm~\ref{alg:emulator}.

\begin{algorithm}
\caption{Stochastic emulator of global climate.}
\label{alg:emulator}
\begin{algorithmic}
\State \textbf{Input:} Temporal evolution of global mean temperature $T_g(t)$
\State \textbf{Output:} Emulated statistics of climate variables

\vspace{1ex}
\State \textbf{Step 1}: Emulate seasonal mean and variance; \\
\hspace{12pt}  \textbullet~Compute seasonal global mean temperature $T_{s,g}$ for each season $s$; \\
\hspace{12pt} \textbullet~For each mode $i$, predict the seasonal mean $\hat{\mu}_{s,i}(T_{s,g})$ and variance $\hat{\sigma}^2_{s,i}(T_{s,g})$.

\State \textbf{Step 2}: Generate stochastic daily fluctuations; \\
\hspace{12pt} \textbullet~At every time step $t$, sample a Gaussian random vector $\boldsymbol{\epsilon} \sim \mathcal{N}(\mathbf{0},\hat{\mathbf{R}})$; \\
\hspace{12pt} \textbullet~Compute the vector autoregressive process $\hat{\boldsymbol{\eta}}(t)$ according to equation (\ref{eq:VAR}).

\State \textbf{Step 3}: Construct time series of spatial fields; \\
\hspace{12pt} \textbullet~Combine seasonal mean and variance with daily fluctuations to obtain $\hat{a}_i(t)$; \\
\hspace{12pt} \textbullet~Multiply PCA coefficients by their mode shapes and superpose all the modes; \\
\hspace{12pt} \textbullet~Denormalize by $\boldsymbol{\sigma}_g$ and $\bar{u}$ to compute $U,V,T,Q$ in physical space (\ref{eq:qhat}).

\State \textbf{Step 4}: Estimate statistics of interest; \\
\hspace{12pt} \textbullet~Average the spatial fields of $U,V,T,Q$ over window to calculate statistics; \\
\hspace{12pt} \textbullet~If needed, input $T_g(t)$ from a different ensemble member, repeat steps 1-3 and average over multiple members.
\end{algorithmic}
\end{algorithm}

\subsection{Nudging the Stochastic Emulator}
\label{app:sde_emulator}

The stochastic emulator introduced previously was designed to capture the second-order statistics of the leading PCA modes. While this emulator has demonstrated effectiveness in representing the conditional Gaussian distribution of certain variables, such as temperature \cite{wang2023emulator}, it inherently struggles to reproduce the non-Gaussian characteristics of climate data, including extreme events associated with higher-order PCA modes.
A common approach to addressing this limitation involves using machine-learning models to debias the emulator.
However, due to the stochastic nature of the emulated spatiotemporal data, instantaneous matches with reference data are not achievable. For instance, an emulated wind speed field on January 1st, 2025, would significantly differ from the corresponding ground truth dataset, whether sourced from ERA5 or CMIP6.
Ideally, an infinite ensemble of realizations could be produced by the emulator, enabling selection of instances closest to reference observations for training a debiasing model.
However, this method is impractical.
A more realistic alternative is the nudging approach \citep{storch_spectral_2000,miguez-macho_regional_2005,sun_impact_2019,huang_development_2021}, where the emulator is forced by the deviation from the reference data to produce a time series of fields that approximately maintain the emulator's statistical characteristics while closely aligning with the observed ground truth.
Herein we interpret how to nudge the stochastic emulator.
In the following, we detail how to implement nudging within the stochastic emulator framework.

Recall the formulation of the emulator (\ref{eq:qhat},\ref{eq:model_ahat}),
\begin{align}
    \hat{a}_{s,i}(t) &= \hat{\mu}_{s,i}\left(T_{s,g} \right) + \hat{\sigma}_{s,i}\left(T_{s,g} \right) \hat{\eta}_{s,i}(t) \\
    \hat{q}_k(\boldsymbol{x},t) &= \bar{u}_k(\boldsymbol{x},t) + \sigma_{g,k} \sum_{i=1}^{500} \hat{a}_i(t) \phi_k^{(i)}(\boldsymbol{x}).
\end{align}
The only stochastic component is the time series of daily fluctuations $\hat{\eta}_{s,i}$.
All other parts are deterministic and constructed to align with the reference data.
Therefore we focus on nudging $\hat{\eta}_{s,i}$, given the true fluctuations $\eta_{s,i}$. Hereafter we omit any subscripts to simplify the notation.

The nudged emulator, denoted as $\boldsymbol{\nu}$, is designed to follow the dynamics of the free-running emulator $\hat{\boldsymbol{\eta}}$, while driven by the deviation from the reference data,
\begin{equation}
    \label{eq:nudged_dt}
    \dot{\boldsymbol{\nu}} = \dot{\hat{\boldsymbol{\eta}}} - \frac{1}{\tau} \left(\boldsymbol{\nu} -  \boldsymbol{\eta}\right).
\end{equation}
The relaxation time scale $\tau$ is a constant that is independent from the season or the PCA mode.
Equation (\ref{eq:nudged_dt}) has a closed-form solution,
\begin{equation}
    \label{eq:nudged_emulator}
    \boldsymbol{\nu}(t) = \boldsymbol{\nu}(0) e^{-t/\tau} + \int_0^t e^{-(t-s)/\tau} \left(\dot{\hat{\boldsymbol{\eta}}}(s) + \frac{1}{\tau} \boldsymbol{\eta}(s)\right) ds.
\end{equation}
The time derivative term $\dot{\hat{\boldsymbol{\eta}}}$ is approximated using the first-order Euler scheme and computed from the free-running emulator data.
Combining the nudged time series of daily fluctuations $\boldsymbol{\nu}$ with seasonal mean and variance, we obtain the complete nudged PCA time series and the spatiotemporal fields,
\begin{align}
    \hat{a}_{s,i}^{\nu}(t) &= \hat{\mu}_{s,i}\left(T_{s,g} \right) + \hat{\sigma}_{s,i}\left(T_{s,g} \right) \hat{\nu}_{s,i}(t) \\
    \hat{q}^{\nu}_k(\boldsymbol{x},t) &= \bar{u}_k(\boldsymbol{x},t) + \sigma_{g,k} \sum_{i=1}^{500} \hat{a}^{\nu}_i(t) \phi_k^{(i)}(\boldsymbol{x}).
\end{align}

The interpretation of nudging and the selection of $\tau$ have been thoroughly discussed in \cite{barthel_sorensen_probabilistic_2024}. Briefly speaking, the relaxation timescale $\tau$ serves to separate the time scales between slow and fast dynamics.
The feedback term in equation (\ref{eq:nudged_dt}) drives the slow dynamics of $\boldsymbol{\nu}$ towards the reference trajectory $\boldsymbol{\eta}$ in the state space, while allowing the fast dynamics of $\boldsymbol{\nu}$ to freely evolve.
Thus, when pairs of the nudged and reference data are used for training a machine-learning model, we are essentially learning a map that corrects the fast features of the imperfect emulator and improve the performance on extreme events.
In our case, the relaxation timescale is set as $\tau = 6 \textrm{hrs}$, consistent with previous work \cite{barthel_sorensen_non-intrusive_2024}. Minor adjustments of $\tau$, such as to 3 or 12 hours, do not significantly alter the results.

The feedback term in (\ref{eq:nudged_dt}), although driving the nudged emulator towards the reference, introduces artificial dissipation not present in the free-running emulator.
Such an effect leads to a distribution of $\boldsymbol{nu}$ and $\hat{\boldsymbol{q}}^{\nu}$ that is slightly different from the free-running emulator. In order for a neural network trained on the nudged dataset to generalize to unseen free-running emulator data, this discrepancy must be remedied.
To this end, we rescale the nudged solution $\hat{\boldsymbol{q}}^{\nu}$ in each season so that its mean and variance match those of the free-running emulator $\hat{\boldsymbol{q}}$ at each grid point.

\section{Machine Learned Debiasing}
\subsection{Conditional Score-based Diffusion model}

Here we describe the training strategy and network architecture used in the ML correction step of our model.
Our model relies on the framework introduced by \citet{barthel_sorensen_non-intrusive_2024}, which aims to learn a deterministic map from the nudged trajectory to the reference trajectory,
\begin{equation}
    \label{eq:map_deterministic}
    \boldsymbol{u} = \mathcal{F}\left(\hat{\boldsymbol{q}}^{\nu} \right).
\end{equation}
In practice, such a mapping is not necessarily deterministic. There could exist multiple reference state $\boldsymbol{u}$ that are close to the same nudged state $\hat{\boldsymbol{q}}^{\nu}$.
Therefore, we generalize this framework by learning a conditional probability distribution function,
\begin{equation}
    \label{eq:map_pdf}
    p(\boldsymbol{u} \mid \hat{\boldsymbol{q}}^{\nu}).
\end{equation}
If the mapping is actually deterministic, the conditional probability distribution will collapse to a Dirac delta function $\delta(\boldsymbol{u} - \mathcal{F}\left(\hat{\boldsymbol{q}}^{\nu} \right))$.
Once the conditional PDF (\ref{eq:map_pdf}) is learned, we can provide the free-running emulation $\hat{\boldsymbol{q}}$ as the conditional information to generate debiased estimations of the state variables $\hat{\boldsymbol{u}}$,
\begin{equation}
    \hat{\boldsymbol{u}}(\boldsymbol{x},t) \sim \mathcal{G}_{\theta,2}\left[\hat{\boldsymbol{q}}(\boldsymbol{x},t) \right]
\end{equation}

Although learning and sampling high-dimensional PDFs were long considered intractable, these tasks have recently become practical thanks to advances in deep generative models.
In this study, we adopt conditional score-based diffusion model \citep{song_score-based_2021,batzolis2021conditional} that has been demonstrated effective for geophysical datasets \citep{bischoff_unpaired_2023}.
Other frameworks, such as flow matching \citep{lipman2022flow} and stochastic interpolant \citep{albergo2023stochastic}, could likewise address the debiasing problem considered here. The choice of the generative model is beyond the scope of this work and will be investigated in the future.

Our implementation of score-based diffusion model follows that of \citet{bischoff_unpaired_2023}.
To simplify the notation, we will use $\boldsymbol{q}$ to represent the nudged emulation $\hat{\boldsymbol{q}}^{\nu}$.
The diffusion model consists of a forward diffusion process, which maps the data distribution to normal distribution, and a reverse denoising process that transforms Gaussian noise to a sample or image of the climate state.
Specifically, given an initial condition $\mathbf{u}(\mathfrak{t} = 0) \sim p_{\mathrm{data}} (\boldsymbol{u} | \boldsymbol{q})$ drawn from the training data, the forward diffusion process is defined by the stochastic differential equation (SDE), 
\begin{equation}
    \label{eq:diffusion_forward}
    \mathrm{d}\mathbf{u} = g(\mathfrak{t}) \mathrm{d\mathbf{W}},
\end{equation}
where the diffusion coefficient $g(\mathfrak{t})$ is a non-negative prescribed function and $\mathrm{\mathbf{W}}$ is a Wienner process.
Note that the diffusion time $\mathfrak{t}$ is within $[0,1]$ and should be distinguished from the physical time $t$.
At any time $\mathfrak{t}$, the solution to the SDE (\ref{eq:diffusion_forward}) is a ``noised'' image $\mathbf{u}(\mathfrak{t})$, which follows a normal distribution conditioned on $\mathbf{u}(0)$,
\begin{equation}
    \label{eq:diffusion_forward_u}
    \mathbf{u}(t) \sim \mathcal{N}\left(\mathbf{u}(0) ,\sigma^2(\mathfrak{t})\right) = p(\mathbf{u}(\mathfrak{t}) | \mathbf{u}(0)),
\end{equation}
where the variance $\sigma^2(\mathfrak{t)}$ depends on $g(\mathfrak{t})$,
\begin{equation}
    \sigma^2(\mathfrak{t}) = \int_0^{\mathfrak{t}} g^2(\mathfrak{t}^{\prime}) \mathrm{d}\mathfrak{t}^{\prime}.
\end{equation}
The marginal distribution of $\mathbf{u}(\mathfrak{t})$ after integrating out $\mathbf{u}(0)$ is defined as $p_{\mathfrak{t}}(\mathbf{u}(\mathfrak{t}) | \mathbf{q})$, which is generally non-Gaussian.
The diffusion coefficient $g(\mathfrak{t})$ is chosen such that at $\mathfrak{t} = 1$, the variance $\sigma^2(\mathfrak{t} = 1)$ has much larger magnitude than the original $\mathbf{u}(\mathfrak{t} = 0)$.
Therefore, $\mathbf{u}(\mathfrak{t} = 0)$ has lost all memory of initial condition, and
\begin{equation}
    \label{eq:diffusion_p_t1}
    p(\mathbf{u}(1) | \mathbf{u}(0)) = p(\mathbf{u}(1)) = \mathcal{N}(\mathbf{0},\sigma^2(1)).
\end{equation}

According to Anderson's theorem \cite{anderson1982reverse}, the reverse of equation (\ref{eq:diffusion_forward}) is also a diffusion process, running backward in time and governed by the following SDE,
\begin{equation}
    \label{eq:diffusion_backward}
    \mathrm{d}\mathbf{u} = -g^2(\mathfrak{t}) \mathbf{s}(\mathbf{u},\mathbf{q},\mathfrak{t}) \mathrm{d}\mathfrak{t} + g(\mathfrak{t})\mathrm{d\overline{\mathbf{W}}},
\end{equation}
where $\overline{\mathbf{W}}$ is a reverse-time Wienner process, and $\mathbf{s}(\mathbf{u},\mathfrak{t})$ is the conditional score,
\begin{equation}
    \label{eq:score}
    \mathbf{s}(\mathbf{u},\mathbf{q}, \mathfrak{t}) = \nabla_{\mathbf{u}} \log p_{\mathfrak{t}} (\mathbf{u}(\mathfrak{t}) | \mathbf{q}). 
\end{equation}
If we have access to the score for all $\mathfrak{t}$, we can derive the reverse diffusion process, simulate it from $\mathfrak{t} = 1$ to $\mathfrak{t} = 0$, and generate samples that follow the data distribution.
To this end, we approximate the score by a neural network, $\mathbf{s}_{\theta}(\mathbf{u},\mathbf{q},\mathfrak{t})$, which is obtained by minimizing the score-matching loss or Fisher's divergence,
\begin{equation}
    \label{eq:loss_SM}
    \mathcal{L}_{S M}(\theta):=\frac{1}{2} \mathbb{E}_{
    \begin{subarray}
        \mathfrak{t} \sim U(0, 1) \\
        \mathbf{u}(\mathfrak{t}),\mathbf{q} \sim p_{\mathfrak{t}}(\mathbf{u}(\mathfrak{t}) | \mathbf{q})
    \end{subarray}
    }
    \left[\sigma^2(\mathfrak{t})\left\|\nabla_{\mathbf{u}} \log p_{\mathfrak{t}}\left( \mathbf{u} | \mathbf{q}\right)-\mathbf{s}_{\theta}(\mathbf{u},\mathbf{q},\mathfrak{t})\right\|_2^2\right],
\end{equation}
where $U(0, 1)$ stands for a uniform distribution from 0 to 1.
However, the loss function (\ref{eq:loss_SM}) cannot be directly optimized, since the true conditional score $\nabla_{\mathbf{u}} \log p_{\mathfrak{t}}\left( \mathbf{u} | \mathbf{q}\right)$ is unknown.
Taking advantage of the Gaussian property of the forward diffusion process (\ref{eq:diffusion_forward_u},\ref{eq:diffusion_p_t1}), \cite{song_score-based_2021} showed that $\mathcal{L}_{SM}$ is equal to the following loss up to an additive term, 
\begin{equation}
    \label{eq:loss}
    \mathcal{L}(\theta) :=\frac{1}{2} \mathbb{E}_{
    \begin{subarray}
        \mathfrak{t} \sim U(0, 1) \\
        \mathbf{u}(0),\mathbf{q} \sim p_(\mathbf{u}_0 | \mathbf{q}) \\
        \mathbf{u}(\mathfrak{t}) \sim p(\mathbf{u}(\mathfrak{t}) | \mathbf{u}(0))
    \end{subarray}
    }
    \left[\sigma^2(\mathfrak{t})\left\|\nabla_{\mathbf{u}} \log p_{\mathfrak{t}}\left( \mathbf{u}(\mathfrak{t}) | \mathbf{u}(0) \right)-\mathbf{s}_{\theta}(\mathbf{u},\mathbf{q},\mathfrak{t})\right\|_2^2\right].
\end{equation}
This expression only involves $\nabla_{\mathbf{u}} \log p_{\mathfrak{t}}\left( \mathbf{u}(\mathfrak{t}) | \mathbf{u}(0) \right)$ which can be computed analytically from the forward diffusion process.

Once the conditional score is learned through training, it can be substituted into the backward SDE (equation \ref{eq:diffusion_backward}) to generate a debiased sample. The backward SDE is simulated using Euler-Maruyama scheme. The complete procedures of training and sampling are summarized in Algorithm \ref{alg:diffusion}.


\begin{algorithm}
\caption{Conditional score-based diffusion model.}
\label{alg:diffusion}
\algblock[Name]{Training}{End}
\begin{algorithmic}
\Training
\State \textbf{Input:} Reference data and nudged emulation $\{\mathbf{u}_i, \mathbf{q}_i\} \sim p(\mathbf{u} | \mathbf{q})$
\Repeat
\State $\mathbf{u}(0),\mathbf{q} \sim p(\mathbf{u} | \mathbf{q})$
\State $\mathfrak{t} \sim U(0,1)$
\State $\mathbf{u}(\mathfrak{t}) \sim p(\mathbf{u}(\mathfrak{t}) | \mathbf{u}(0)$
\State Take gradient descent step on $\nabla_{\theta}\left[\sigma^2(\mathfrak{t})\left\|\nabla_{\mathbf{u}} \log p_{\mathfrak{t}}\left( \mathbf{u}(\mathfrak{t}) | \mathbf{u}(0) \right)-\mathbf{s}_{\theta}(\mathbf{u}, \mathbf{q}, \mathfrak{t})\right\|_2^2\right]$
\Until{converged}
\State \textbf{Output:} Trained neural network $\mathbf{s}_{\theta}(\mathbf{u},\mathbf{q},\mathfrak{t})$.
\End
\end{algorithmic}
\algblock[Name]{Sampling}{End}
\begin{algorithmic}
\Sampling
\State \textbf{Input:} Snapshot of emulated state $\mathbf{q}$
\State $\mathbf{u}(1) \sim \mathcal{N}(\mathbf{0},\sigma^2(1))$
\For{$\mathfrak{t} = 1$ \textbf{to} 0}
    \State Evaluate $\mathbf{s}_{\theta}(\mathbf{u}(\mathfrak{t}),\mathbf{q},\mathfrak{t})$
    \State $\mathbf{u}(\mathfrak{t} - \Delta \mathfrak{t}) = \mathbf{u}(\mathfrak{t}) + g^2(\mathfrak{t}) \mathbf{s}_{\theta}(\mathbf{u},\mathbf{q},\mathfrak{t}) \Delta \mathfrak{t} - g(\mathfrak{t}) \left(\overline{\mathbf{W}}(\mathfrak{t}) - \overline{\mathbf{W}}(\mathfrak{t} - \Delta \mathfrak{t}) \right)$
    \State $\mathfrak{t} \gets \mathfrak{t} - \Delta \mathfrak{t}$
\EndFor
\State \textbf{Output:} Debiased snapshot $\hat{\mathbf{u}} = \mathbf{u}(\mathfrak{t} = 0)$
\End
\end{algorithmic}
\end{algorithm}

\subsection{Network Architecture and Training Parameters}
Before feeding the emulation and reference data into the diffusion model, we remove the true climatological mean and scale each variable ($U,V,T,Q$) by twice its own globally-averaged standard deviation.
In other words, we focus on correcting the fluctuation fields provided by the emulator, and each variable is scaled to the same order of magnitude.
Regarding the diffusion coefficient $g(\mathfrak{t})$ in equation (\ref{eq:diffusion_forward}), we adopt the ``variance-exploding'' schedule,
\begin{equation}
    g(\mathfrak{t}) = \sigma_{\min} \left(\frac{\sigma_{\max}}{\sigma_{\min}} \right)^{\mathfrak{t}} \sqrt{2 \log \left(\frac{\sigma_{\max}}{\sigma_{\min}} \right) },
\end{equation}
where $\sigma_{\min} = 0.01$ and $\sigma_{\max}$ is chosen as the maximum 2-norm distance between any two snapshots of $\mathbf{u}$.

The neural network architecture we adopted is a U-Net \cite{ronneberger_u-net_2015,bischoff_unpaired_2023}, denoted as
\begin{equation}
    \label{eq:UNet}
    \mathbf{s}_{\theta}(\mathbf{u},\mathbf{q},\mathfrak{t}) = \mathcal{U}(\mathbf{X},\mathfrak{t}; \boldsymbol{\theta}).
\end{equation}
The first input $X$ is a tensor of size $(N_1,N_2,C_{in})$, where $(N_1,N_2)$ are longitude and latitude dimensions, and $C_{in}$ is the number of channels.
In our case, $C_{in} = 8$, including $(U,V,T,Q)$ from the nudged emulation and reference data, respectively.
The output of our U-Net is another tensor of size $(N_1,N_2,C_{out})$.
The number of output channels is $C_{out} = 4$.
The U-Net architecture consists of
\begin{enumerate}
    \item A lifting layer which increases the number of channels form $C_{in}$ to 32;
    \item Three downsampling convolutional layers, each of which reduces the spatial dimension and increase the number of channels by a factor of 2;
    \item Eight residual blocks \citep{he_deep_2015} to promote continuity in the latent space;
    \item Three nearest neighbor up sampling layer and convolution layers which mirror the downsampling operations;
    \item A Final projection layer that decreases the number of channels to $C_{out}$.  
\end{enumerate}

Our choice of the optimizer follows \cite{song_score-based_2021} and \cite{bischoff_unpaired_2023}.
An Adam optimizer is adopted with a learning rate of $\lambda_0 = 2e-4$, $\epsilon = 1e-8$, $\beta_1 = 0.9$, $\beta_2 = 0.999$. The gradient norm clipping is employed to a value of 1.0. For both the ERA5 and CMIP6 datasets, we set the batchsize as 8, and train the U-Net for 200 epochs.



\section{Data Post-processing and Evaluation Metrics}
\label{app:post}
This section provides detailed definitions and calculation methods for all statistics and metrics presented in the main figures.
Given that the climatological mean $\bar{\boldsymbol{u}}(\boldsymbol{x},t)$ (equation \ref{eq:q_mean}) is assumed known, our analysis focuses on evaluating the statistics of the fluctuation fields, $\boldsymbol{u} - \bar{\boldsymbol{u}}$.
This approach enables a clearer comparison between reference statistics and GEN$^2$ prediction.
In the following subsections, the fluctuations from the climatological mean, $\boldsymbol{u} - \bar{\boldsymbol{u}}$, will be simply written as $\boldsymbol{u}$ for notational convenience.

\subsection{Single-point and two-point statistics}
Without loss of generality, we use the zonal wind speed $U(\boldsymbol{x},t)$ as an example.
To evaluate the statistics of $U$ at location $\boldsymbol{x}$, we perform a time average (e.g. from 1979 to 2018 for ERA5 dataset).
If we have $N_t$ time steps available, the mean and standard deviation are computed as,
\begin{equation}
    \mu_U(\boldsymbol{x},t_j) = \frac{1}{N_t} \sum_{j = 1}^{N_t} U(\boldsymbol{x},t_j)
    \quad \textrm{and} \quad
    \sigma_U(\boldsymbol{x}) = \sqrt{\frac{1}{N_t - 1}  \sum_{j = 1}^{N_t} \left(U(\boldsymbol{x},t_j) - \mu_U(\boldsymbol{x},t_j) \right)^2}.
\end{equation}
To obtain the unbiased skewness, we first compute
\begin{equation}
    s_U(\boldsymbol{x}) = \frac{\frac{1}{N_t} \sum_{j = 1}^{N_t} \left(U(\boldsymbol{x},t_j) - \mu_U(\boldsymbol{x},t_j) \right)^3 }{\left( \frac{1}{N_t} \sum_{j = 1}^{N_t} \left(U(\boldsymbol{x},t_j) - \mu_U(\boldsymbol{x},t_j) \right)^2 \right)^{3/2}},
\end{equation}
which is then substituted into
\begin{equation}
    s^0_U(\boldsymbol{x}) = \frac{\sqrt{N_t (N_t - 1)}}{N_t - 2} s_U(\boldsymbol{x}).
\end{equation}
The calculation of unbiased kurtosis also consists of two steps:
\begin{equation}
    k_U(\boldsymbol{x}) = \frac{\frac{1}{N_t} \sum_{j = 1}^{N_t} \left(U(\boldsymbol{x},t_j) - \mu_U(\boldsymbol{x},t_j) \right)^4 }{\left( \frac{1}{N_t} \sum_{j = 1}^{N_t} \left(U(\boldsymbol{x},t_j) - \mu_U(\boldsymbol{x},t_j) \right)^2 \right)^{2}},
\end{equation}
\begin{equation}
    k^0_U(\boldsymbol{x}) = \frac{N_t - 1}{(N_t - 2)(N_t - 3)} \left((N_t + 1)k_U(\boldsymbol{x}) - 3(N_t - 1) + 3\right).
\end{equation}

At the same location $\boldsymbol{x}$, the correlation coefficient between two variables $U$ and $V$ are defined as,
\begin{equation}
    \rho(U,V) = \frac{\mathrm{cov}\left(U(\boldsymbol{x},t),V(\boldsymbol{x},t) \right)}{\mathrm{cov}\left(U(\boldsymbol{x},t),U(\boldsymbol{x},t) \right)\mathrm{cov}\left(V(\boldsymbol{x},t),V(\boldsymbol{x},t) \right)},
\end{equation}
where $\mathrm{cov}\left(U(\boldsymbol{x},t),V(\boldsymbol{x},t)\right)$ is the time-averaged covariance between $U$ and $V$.

The two-point correlation coefficient of $U$ is defined as,
\begin{equation}
    \rho(U(\boldsymbol{x}_0),U(\boldsymbol{x})) = \frac{\mathrm{cov}\left(U(\boldsymbol{x}_0,t),U(\boldsymbol{x},t) \right)}{\mathrm{cov}\left(U(\boldsymbol{x}_0,t),U(\boldsymbol{x}_0,t) \right)\mathrm{cov}\left(U(\boldsymbol{x},t),U(\boldsymbol{x},t) \right)}.
\end{equation}
The anchor point $\boldsymbol{x}_0$ is selected as major cities (e.g. Boston, Hong Kong) in figure 3, 6 and in section \ref{app:results}.

The global root-mean-square error (RMSE) of an arbitrary statistic $\mathcal{Q}$ (e.g. in figure 5 and table \ref{table:cov_1},\ref{table:cov_2}) is defined as,
\begin{equation}
    \label{eq:RMSE_g}
    \mathrm{RMSE}(\mathcal{Q}) = \left[ \frac{1}{S}\int_S \left(\hat{\mathcal{Q}}(\theta,\varphi,t) - \mathcal{Q} (\theta,\varphi,t)\right)^2 \cos\theta d\theta d\varphi\right]^{1/2},
\end{equation}
where $\hat{\mathcal{Q}}$ is the GEN$^2$ prediction and $\mathcal{Q}$ is the reference statistics.

\subsection{Wheeler-Kiladis spectrum}
\begin{figure}
    \centering
    \includegraphics[width=0.8\linewidth]{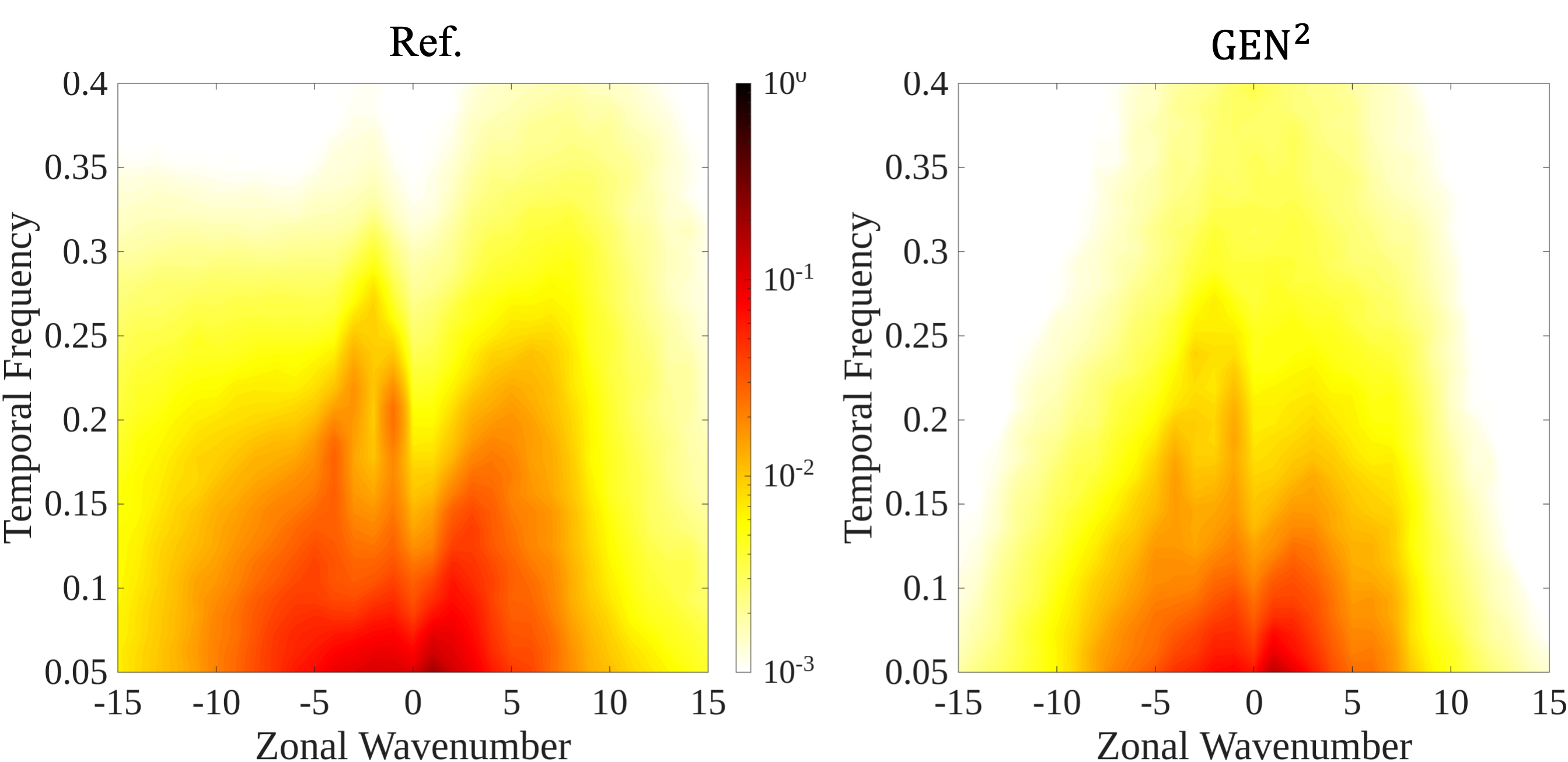}
    \caption{Raw Wheeler-Kiladis spectrum of zonal wind, computed using 1979-2018 ERA5 reference data and GEN$^2$ prediction. These spectra are normalized by the ``background power'' to obtain the spectra in figure 2(c) of the main manuscript.}
    \label{fig:era5_kiladis_raw}
\end{figure}

The Wheeler-Kiladis spectrum is computed following the procedure described in \citet{wheeler_convectively_1999,kiladis_convectively_2009}. Given data as a function of longitude, latitude, and time $[(\theta,\phi,t)$ the latitude range is first truncated to $\phi \in [- 15^\circ,15^\circ]$. Then for each latitude $\phi_j$, and time $t_j$ the Fourier spectrum is computed in the azimuthal direction $\theta$ giving rise to a azimuthal wavenumber $m$. Then for each $\phi_j$ and $m_j$ the time series of data is split into a series of overlapping segments. Following \cite{wheeler_convectively_1999,kiladis_convectively_2009} we set the length of each segment to 96 days and the overlap to 65 days. Each segment is then detrended using a linear fit and Fourier transformed in time - giving a temporal frequency $f$.
After averaging over all latitudes and all temporal segments, the raw Wheeler-Kiladis spectra are shown in figure \ref{fig:era5_kiladis_raw}.

Due to the strong ``redness'' of the spectra, detailed features corresponding to the equatorial waves are obscured. To better identify the ridges of the spectra, we first apply a 1-2-1 filter ten times to obtain a much smoother ``background'' spectra.
Then the raw spectra in figure \ref{fig:era5_kiladis_raw} are divided by the background \cite{wheeler_convectively_1999}. The results, as shown in figure 2(c) of the main manuscript, more clearly show the spectral peaks that correspond to different types of equatorial waves.
Note that the spectra in our results should not be directly compared against the plots in \cite{wheeler_convectively_1999}.
The reason is that their analysis was based on the long-wave radiation data, which are proxy for cloudiness, whereas our analysis focuses on near-surface zonal wind speed.

\section{Additional Results: Bias Reduction via ML Correction}
\label{app:results}

\begin{figure}
    \centering
    \includegraphics[width=0.95\textwidth]{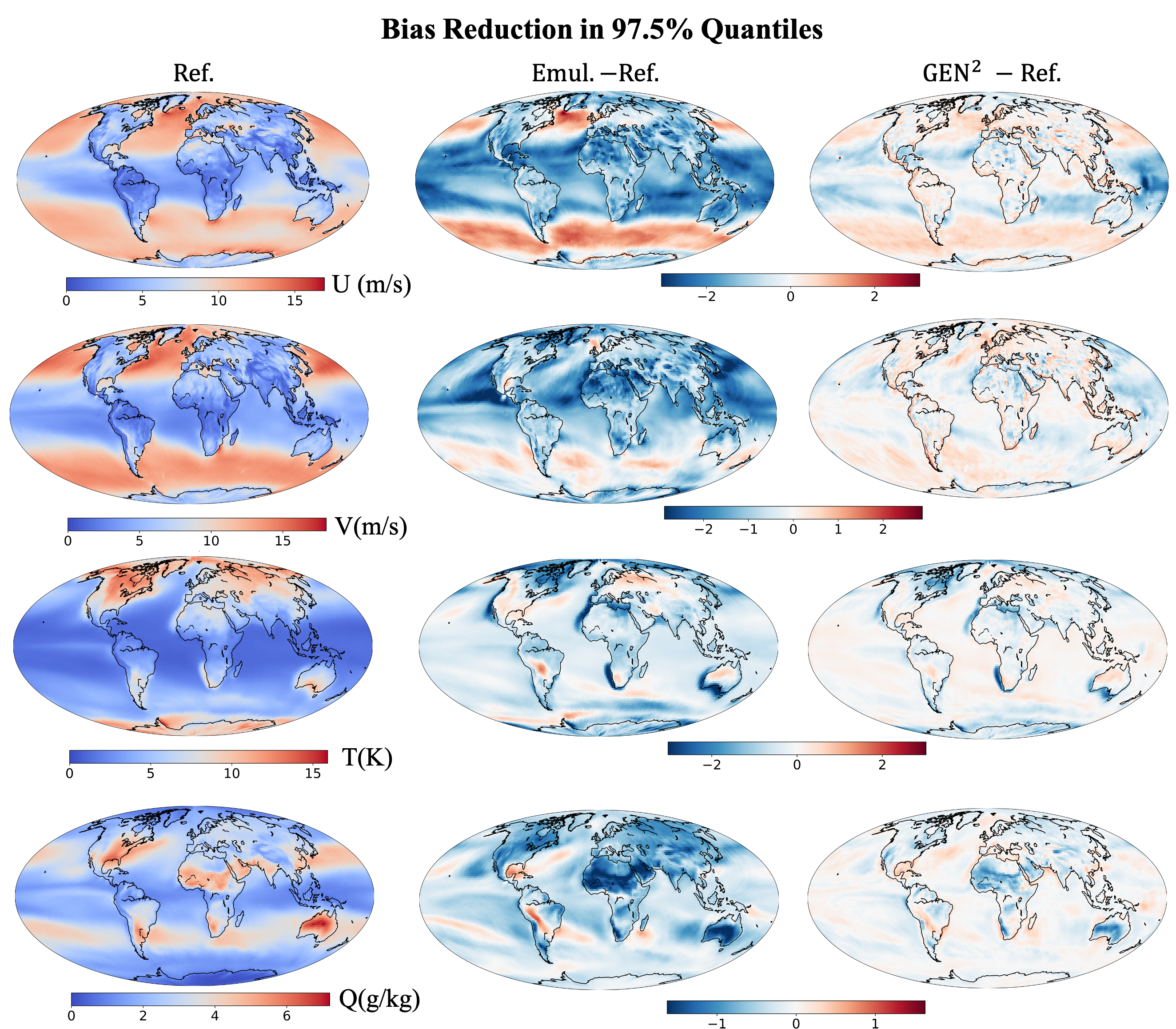}
    \caption{Left column: 97.5\% quantile of zonal wind, meridional wind, temperature, computed from reference data.
    Middle column: bias of conditional Gaussian emulation.
    Right column: bias of ``GEN$^2$'' approach.}
    \label{fig:app_ml_quantile}
\end{figure}

To illustrate the debiasing capabilities of the ML correction step, we show in figure \ref{fig:app_ml_quantile} the bias in the 97.5\% quantile, predicted with or without ML correction.
As explained at the beginning of Appendix \ref{app:post}, the statistics are evaluated for the fluctuation fields.
Before applying ML correction (middle column), the error of the conditional Gaussian emulator is already moderately accurate.
For example, the error of $T$ at most locations is within 3$K$, and the highest error is within 3$K$. 
The ML model (right column) consistently reduces the error of all the state variables at almost all the locations.
A more quantitative comparison is provided in table \ref{table:qntl_1},\ref{table:qntl_2}.
The bias in standard deviation, quantile, skewness, and kurtosis are all significantly reduced by ML correction.

\begin{figure}
    \centering
    \includegraphics[width=0.95\textwidth]{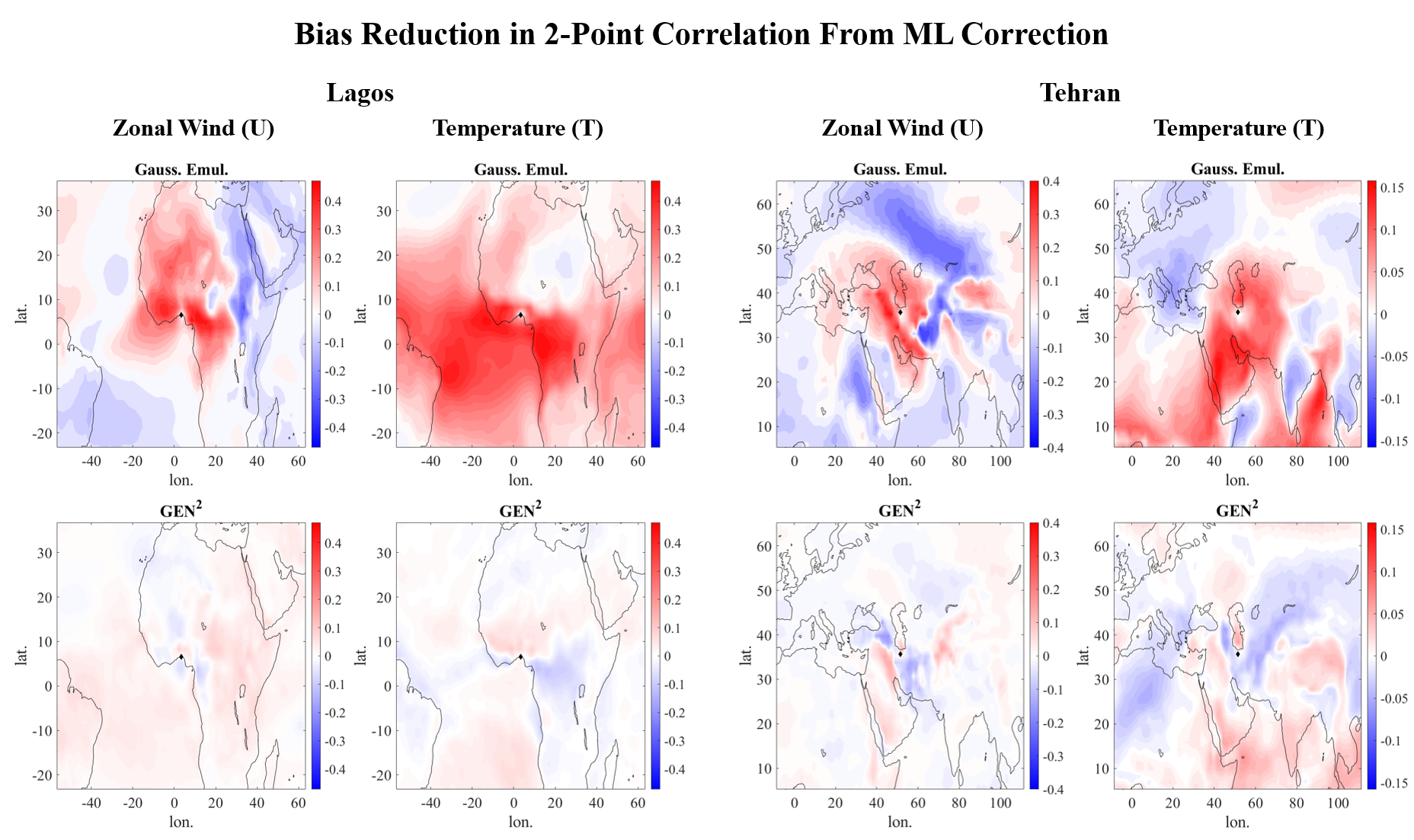}
    \caption{Bias in two-point correlations of zonal wind and temperature, centered at Lagos and Tehran. Contour plots represent bias relative to reference ERA5 data. Panels labeled ``Gauss. Emul.'' correspond to predictions of the conditional Gaussian emulator \textit{only} (no ML correction) and ``GEN$^2$'' represents the full model prediction (with ML correction) .}
    \label{fig:app_ml_2point}
\end{figure}

The bias reduction in two-point correlations are shown in figure \ref{fig:app_ml_2point}.
We select Lagos and Tehran for visualization, because the bias of the conditional Gaussian emulator is more pronounced at these two locations.
Top panels in figure \ref{fig:app_ml_2point} are the bias of the conditional Gaussian emulator, without ML correction.
Bottom panels are the bias of GEN$^2$ prediction.

\begin{table}
\centering
\begin{tabular}{|c| r r r| r r r|}
\hline
Statistics & \multicolumn{3}{|c|}{RMSE of $U$ stats} & \multicolumn{3}{|c|}{RMSE of $V$ stats} \\
 & Em. & GEN$^2$ & Change & Em. & GEN$^2$ & Change \\
\hline
Std & 0.43 & 0.15 & -65\% & 0.47 & 0.13 & -73\%\\
97.5\% quantile & 1.14 & 0.46 & -60\% & 0.99 & 0.33 & -66\%\\
Skewness & 0.41  & 0.19 & -53\% & 0.27  & 0.14  &  -48\% \\
Kurtosis & 0.84 & 0.57  & -33\% & 0.68 & 0.46  &  -33\% \\
\hline
\end{tabular}
\caption{RMSE of single-point statistics of $U$, $V$, defined as equation (\ref{eq:RMSE_g}). Columns labeled as ``Em.'' are the RMSE of the prediction of conditional Gaussian emulator, and ``GEN$^2$'' is the full-model prediction. ``Change'' columns are the relative error change from conditional Gassuain emulator to GEN$^2$, more precisely, $(\mathrm{RMSE}(\mathrm{GEN}^2) - \mathrm{RMSE}(\mathrm{Em.})) / \mathrm{RMSE}(\mathrm{Em.})$.}
\label{table:qntl_1}
\end{table}

\begin{table}
\centering
\begin{tabular}{|c| r r r| r r r|}
\hline
Statistics & \multicolumn{3}{|c|}{RMSE of $T$ stats} & \multicolumn{3}{|c|}{RMSE of $Q$ stats} \\
 & Em. & GEN$^2$ & Change & Em. & GEN$^2$ & Change \\
\hline
Std & 0.23 & 0.10 & -56\% & 0.19 & 0.05 & -72\%\\
97.5\% quantile &  0.67 & 0.35 & -48\% & 0.46 & 0.17 & -63\%\\
Skewness & 0.34 & 0.20 & -42\% & 0.50  & 0.27 & -47\% \\
Kurtosis & 0.89  & 0.68  &  -24\% & 1.78 & 1.38 &  -23\% \\
\hline
\end{tabular}
\caption{Same as table \ref{table:qntl_1}, but for temperature $T$ and $Q$.}
\label{table:qntl_2}
\end{table}

Table \ref{table:cov_1} and \ref{table:cov_2} summarize the error of two-point correlations at more locations, selected from different regions and climate over the world.
At most locations, the conditional Gaussian emulator already achieves an accurate prediction, with RMSE lower than 0.03.
After applying ML correction, the errors are significantly reduced at all the locations considered.
These results demonstrate the robustness of the ML correction.

\begin{table}
\centering

\begin{tabular}{|l r r| r r r| r r r|}
\hline
\multicolumn{3}{|c|}{Anchor point $\boldsymbol{x}_0$} & \multicolumn{3}{|c|}{RMSE of $\rho(U(\boldsymbol{x}_0),U(\boldsymbol{x}))$} & \multicolumn{3}{|c|}{RMSE of $\rho(V(\boldsymbol{x}_0),V(\boldsymbol{x}))$} \\
City & Lon(E) & Lat(N) & Em. & GEN$^2$ & Change & Em. & GEN$^2$ & Change \\
\hline
Boston & -71.1 & 42.4 & 0.016 & 0.009 & -43\% & 0.018 & 0.008 & -53\% \\
Los Angeles & -118.2 & 34.1 & 0.028 & 0.011 & -62\% & 0.026 & 0.009 & -67\% \\
Chicago & -87.6 & 41.9 & 0.020 & 0.008 & -59\% & 0.017 & 0.008 & -54\% \\
Houston & -95.4 & 29.8 & 0.020 & 0.009 & -57\% & 0.017 & 0.009 & -48\% \\
Kansas City & -94.6 & 39.1 & 0.022 & 0.008 & -64\% & 0.017 & 0.008 & -51\% \\
London & -0.1 & 51.5 & 0.018 & 0.009 & -50\% & 0.019 & 0.008 & -60\% \\
Anchorage & -149.9 & 61.2 & 0.019 & 0.012 & -36\% & 0.021 & 0.009 & -58\% \\
Paris & 2.4 & 48.9 & 0.019 & 0.009 & -53\% & 0.019 & 0.008 & -57\% \\
Athens & 23.7 & 38.0 & 0.024 & 0.010 & -57\% & 0.022 & 0.009 & -60\% \\
Moscow & 37.6 & 55.8 & 0.024 & 0.010 & -58\% & 0.022 & 0.008 & -66\% \\
Stockholm & 18.1 & 59.3 & 0.022 & 0.010 & -57\% & 0.020 & 0.008 & -61\% \\
Tokyo & 139.7 & 35.7 & 0.017 & 0.008 & -51\% & 0.017 & 0.009 & -48\% \\
Hong Kong & 114.2 & 22.3 & 0.026 & 0.009 & -64\% & 0.025 & 0.009 & -65\% \\
New Delhi & 77.1 & 28.6 & 0.028 & 0.010 & -64\% & 0.028 & 0.010 & -65\% \\
Tehran & 51.4 & 35.7 & 0.030 & 0.011 & -63\% & 0.034 & 0.010 & -70\% \\
Astana & 71.5 & 51.2 & 0.022 & 0.009 & -60\% & 0.022 & 0.009 & -60\% \\
Cairo & 31.2 & 30.0 & 0.029 & 0.009 & -70\% & 0.027 & 0.008 & -69\% \\
Cape Town & 18.4 & -33.9 & 0.018 & 0.009 & -50\% & 0.022 & 0.009 & -60\% \\
Lagos & 3.4 & 6.5 & 0.043 & 0.015 & -65\% & 0.040 & 0.009 & -78\% \\
Kisangani & 25.2 & 0.1 & 0.047 & 0.015 & -69\% & 0.036 & 0.009 & -76\% \\
Mombasa & 39.7 & -4.0 & 0.041 & 0.018 & -55\% & 0.038 & 0.010 & -74\% \\
Sydney & 151.2 & -33.9 & 0.020 & 0.009 & -56\% & 0.020 & 0.008 & -58\% \\
Brasília & -47.9 & -15.8 & 0.029 & 0.012 & -58\% & 0.044 & 0.010 & -78\% \\
Bogota & -74.1 & 4.7 & 0.054 & 0.023 & -57\% & 0.034 & 0.017 & -49\% \\
Buenos Aires & -58.4 & -34.6 & 0.020 & 0.009 & -55\% & 0.019 & 0.009 & -54\% \\
\hline
\end{tabular}
\caption{RMSE of two-point correlation of $U$, $V$. Columns labeled as ``Em.'' are the RMSE of the prediction of conditional Gaussian emulator, and ``GEN$^2$'' is the full-model prediction. ``Change'' columns are the relative error change from conditional Gassuain emulator to GEN$^2$, more precisely, $(\mathrm{RMSE}(\mathrm{GEN}^2) - \mathrm{RMSE}(\mathrm{Em.})) / \mathrm{RMSE}(\mathrm{Em.})$.}
\label{table:cov_1}
\end{table}

\begin{table}
\centering

\begin{tabular}{|l r r| r r r| r r r|}
\hline
\multicolumn{3}{|c|}{Anchor point $\boldsymbol{x}_0$} & \multicolumn{3}{|c|}{RMSE of $\rho(T(\boldsymbol{x}_0),T(\boldsymbol{x}))$} & \multicolumn{3}{|c|}{RMSE of $\rho(Q(\boldsymbol{x}_0),Q(\boldsymbol{x}))$} \\
City & Lon(E) & Lat(N) & Em. & GEN$^2$ & Change & Em. & GEN$^2$ & Change \\
\hline
Boston & -71.1 & 42.4 & 0.017 & 0.012 & -30\% & 0.016 & 0.009 & -43\% \\
Los Angeles & -118.2 & 34.1 & 0.021 & 0.011 & -48\% & 0.022 & 0.010 & -54\% \\
Chicago & -87.6 & 41.9 & 0.014 & 0.010 & -29\% & 0.017 & 0.009 & -46\% \\
Houston & -95.4 & 29.8 & 0.019 & 0.014 & -28\% & 0.015 & 0.008 & -47\% \\
Kansas City & -94.6 & 39.1 & 0.014 & 0.010 & -30\% & 0.016 & 0.009 & -45\% \\
London & -0.1 & 51.5 & 0.023 & 0.010 & -55\% & 0.022 & 0.011 & -49\% \\
Anchorage & -149.9 & 61.2 & 0.026 & 0.013 & -48\% & 0.026 & 0.014 & -48\% \\
Paris & 2.4 & 48.9 & 0.023 & 0.010 & -54\% & 0.023 & 0.012 & -47\% \\
Athens & 23.7 & 38.0 & 0.018 & 0.012 & -35\% & 0.023 & 0.009 & -60\% \\
Moscow & 37.6 & 55.8 & 0.017 & 0.011 & -36\% & 0.021 & 0.010 & -51\% \\
Stockholm & 18.1 & 59.3 & 0.021 & 0.010 & -53\% & 0.021 & 0.011 & -49\% \\
Tokyo & 139.7 & 35.7 & 0.019 & 0.011 & -40\% & 0.015 & 0.009 & -39\% \\
Hong Kong & 114.2 & 22.3 & 0.023 & 0.014 & -41\% & 0.020 & 0.011 & -46\% \\
New Delhi & 77.1 & 28.6 & 0.031 & 0.018 & -43\% & 0.022 & 0.011 & -47\% \\
Tehran & 51.4 & 35.7 & 0.030 & 0.018 & -42\% & 0.033 & 0.010 & -68\% \\
Astana & 71.5 & 51.2 & 0.017 & 0.011 & -32\% & 0.026 & 0.011 & -59\% \\
Cairo & 31.2 & 30.0 & 0.031 & 0.014 & -55\% & 0.027 & 0.008 & -69\% \\
Cape Town & 18.4 & -33.9 & 0.022 & 0.012 & -45\% & 0.022 & 0.010 & -55\% \\
Lagos & 3.4 & 6.5 & 0.094 & 0.018 & -81\% & 0.034 & 0.013 & -63\% \\
Kisangani & 25.2 & 0.1 & 0.072 & 0.015 & -79\% & 0.046 & 0.011 & -77\% \\
Mombasa & 39.7 & -4.0 & 0.100 & 0.023 & -77\% & 0.059 & 0.015 & -75\% \\
Sydney & 151.2 & -33.9 & 0.025 & 0.012 & -52\% & 0.021 & 0.010 & -53\% \\
Brasília & -47.9 & -15.8 & 0.042 & 0.013 & -68\% & 0.030 & 0.012 & -62\% \\
Bogota & -74.1 & 4.7 & 0.092 & 0.026 & -71\% & 0.054 & 0.025 & -53\% \\
Buenos Aires & -58.4 & -34.6 & 0.019 & 0.012 & -38\% & 0.017 & 0.010 & -42\% \\
\hline
\end{tabular}
\caption{Same as table \ref{table:cov_1}, but for temperature $T$ and humidity $Q$.}
\label{table:cov_2}
\end{table}

\end{appendices}



\bibliography{references_used}

\end{document}